\documentclass[draft]{agujournal2019}
\usepackage{url} 
\usepackage{lineno}
\usepackage[inline]{trackchanges} 
\usepackage{soul}
\usepackage{amssymb}

\draftfalse
\justifying

\journalname{JGR: Space Physics}

\begin{document}

\title{Ion Acceleration at Magnetotail Plasma Jets}

\authors{L.~Richard\affil{1,2}, 
        Yu.~V.~Khotyaintsev\affil{1}, 
        D.~B.~Graham\affil{1}, 
        A.~Vaivads\affil{3}, 
        R.~Nikoukar\affil{4}, 
        I.~J.~Cohen\affil{4}, 
        D.~L.~Turner\affil{4}, 
        S.~A.~Fuselier\affil{5, 6}, 
        C.~T.~Russell\affil{7}}

\affiliation{1}{Swedish Institute of Space Physics, Uppsala, Sweden}
\affiliation{2}{Space and Plasma Physics, Department of Physics and Astronomy, Uppsala University, Sweden}
\affiliation{3}{Space and Plasma Physics, School of Electrical Engineering, KTH Royal Institute of Technology, Stockholm, Sweden}
\affiliation{4}{The Johns Hopkins University Applied Physics Laboratory, Laurel, Maryland, USA}
\affiliation{5}{Southwest Research Institute, San Antonio, TX, USA}
\affiliation{6}{Department of Physics and Astronomy, University of Texas at San Antonio, San Antonio, TX, USA}
\affiliation{7}{Department of Earth Planetary and Space Sciences, University of California Los Angeles, Los Angeles, CA, USA}

\correspondingauthor{Louis Richard}{louis.richard@irfu.se}

\begin{keypoints}
\item Heavy ions of solar wind origin dominate fluxes at energies $> 100$ keV due to charge-state-dependent energization.
\item Ions with gyroradii smaller than the scale of the structure ($\rho_p < L_{y}$) gain energy from the ion bulk flow.
\item Ions with gyroradii $\rho_p \gtrsim L_{y}$ gain energy from non-adiabatic interaction with the electric field pulse.
\end{keypoints}

\begin{abstract}
We investigate a series of Earthward bursty bulk flows (BBFs) observed by the Magnetospheric Multiscale (MMS) spacecraft in Earth’s magnetotail at (-24, 7, 4)~$R_E$ in Geocentric Solar Magnetospheric (GSM) coordinates. At the leading edges of the BBFs, we observe complex magnetic field structures. In particular, we focus on one which presents a chain of small scale ($\sim 0.5~R_E$) dipolarizations, and another with a large scale ($\sim 3.5~R_E$) dipolarization. Although the two structures have different scales, both of these structures are associated with flux increases of supra-thermal ions with energies $\gtrsim 100~\textrm{keV}$. We investigate the ion acceleration mechanism and its dependence on the mass and charge state. We show that the ions with gyroradii smaller than the scale of the structure are accelerated by the ion bulk flow. We show that whereas in the small scale structure, ions with gyroradii comparable with the scale of the structure undergo resonance acceleration, and the acceleration in the larger scale structure is more likely due to a spatially limited electric field.
\end{abstract}

\section{Introduction}
Plasma jets are ubiquitous phenomena in the universe \cite{phan_extended_2000,masuda_loop-top_1994,pudritz_magnetic_2012}. In particular, in Earth's magnetotail, jets take the form of high speed ($V > 400~\textrm{km}~\textrm{s}^{-1}$) transient Earthward plasma flows referred to as bursty bulk flows (BBFs) \cite{angelopoulos_bursty_1992,angelopoulos_statistical_1994}. These BBFs play a crucial role in magnetotail activity \cite{sitnov_explosive_2019}. BBFs are seen as short living narrow flow channels which last 10-100 s \cite{baumjohann_characteristics_1990} and extend over $\sim 1 - 5~R_E$ in the dawn-dusk direction \cite{nakamura_spatial_2004}. At the leading edge, BBFs are accompanied by a sharp dipolarization (i.e., increase of the northward magnetic field $B_z$) of the magnetic field called dipolarization fronts (DFs) \cite{nakamura_motion_2002,ohtani_temporal_2004,runov_themis_2009,sergeev_kinetic_2009}. 

DFs are thin ion-scale boundaries similar to vertical current sheets \cite{runov_themis_2011,liu_current_2013}, which separate the lower temperature dense plasma ahead of the DF from the hotter tenuous plasma behind the DF \cite{khotyaintsev_plasma_2011,fu_energetic_2013}. Various mechanisms are thought to be responsible for the formation of DFs, such as the ballooning/interchange instability \cite{pritchett_kinetic_2010} and unsteady magnetic reconnection \cite{sitnov_dipolarization_2009,fu_energetic_2013}.

Electron acceleration associated with DFs has been widely observed using Cluster \cite{apatenkov_multi-spacecraft_2007}, THEMIS \cite{runov_themis_2009} and MMS \cite{turner_energy_2016}. The two main mechanisms responsible for the acceleration of electrons are the betatron acceleration due to Earthward transport of electrons in the increasing magnetic field \cite{birn_particle_2013} and Fermi acceleration due to shrinking of the magnetic field lines during the Earthward convection \cite{fu_fermi_2011}. In particular,  \citeA{fu_fermi_2011} showed that betatron acceleration dominates in the growing flux pileup region, while Fermi acceleration dominates in the decaying flux pileup region. On the other hand, \citeA{malykhin_contrasting_2018} showed that changes in the spectral index of electrons is sometimes associated with high-frequency waves. \citeA{malykhin_contrasting_2018} suggested that the spectral index changes are due to non-adiabatic wave-particle interaction, which results in electron acceleration and/or scattering.

Ion acceleration at DFs has been extensively studied by means of numerical simulations and observations \cite{fu_magnetotail_2020}. In particular, test particle simulations in assumed or simulated electromagnetic field pulses carried out by \citeA{birn_particle_1994} and \citeA{birn_substorm_1997,birn_particle_2000,birn_particle_2013,birn_energetic_2015,birn_ion_2017}, showed that the ions get accelerated by the motional cross-tail electric field, while gyrating inside the $\mathbf{E} \times \mathbf{B}$ Earthward-convecting structure. Since the magnetic field increased during the Earthward convection, they concluded that the ions undergo a quasi-adiabatic betatron-like acceleration. On the other hand, observations and the particle simulation by \citeA{zhou_accelerated_2010,zhou_nature_2011}, showed the formation of an energized ion population ahead of the DF due to ion reflection from the DF similar to that at shocks \cite{gosling_evidence_1982}. \citeA{zhou_nature_2011} showed that in the presence of a finite northward magnetic field, the energized population is confined to an ion-scale region ahead of the DF. Similarly, \citeA{ukhorskiy_rapid_2013} showed that the negative magnetic dip observed in the soliton-like magnetic field structure \cite{runov_themis_2009} is bounded by two magnetic null points, which are local maximum and minimum of the effective potential energy, and hence creates a potential well. The potential well leads to trapping of the ions between the DF (local potential maximum) and the reconnection point (local potential minimum) ahead of the DF. Recently, \citeA{ukhorskiy_ion_2017,ukhorskiy_ion_2018} showed that ions can be stably trapped in the inverse magnetic field gradient and gain energy through persistent interactions with the motional electric field.

Moreover, \citeA{ukhorskiy_ion_2017,ukhorskiy_ion_2018} showed that the acceleration enabled by stable trapping in the inverse magnetic field gradient is charge-state-dependent and mass independent. Similarly, the test particle simulation of ion acceleration in stochastic electromagnetic perturbations carried out by \citeA{catapano_charge_2017}, showed that the energization is linearly dependent on the charge state, while only weakly mass dependant. \citeA{mitchell_energetic_2018} showed using the Van Allen Probes at $5 < L < 7$ and a simple model of Earthward flow limited in azimuthal extent, that the energization is adiabatic or quasi-adiabatic and therefore, the particle trajectories and energization depends only on ion charge state and not on the ion mass. Using, the Energetic Ion Spectrometer (EIS) on the Magnetospheric Multiscale (MMS) spacecraft, \citeA{bingham_chargestatedependent_2020} found a case of ion energization in the magnetotail ordered by charge state.

Nevertheless, due to their larger gyroradii, ion acceleration is much more complicated than electron acceleration, so that, detailed observations of ion acceleration at DFs are still lacking with respect to those of electron acceleration. In particular, the identification of the different acceleration mechanisms acting at scales smaller than, comparable to, and larger than the scale of the magnetic field structures and the adiabaticity of the acceleration mechanism at the aforementioned scales are still open questions. Here, we investigate the ion acceleration at turbulent reconnection jet fronts in order to advance our understanding of these questions.

\section{Observations}\label{sec:observations}
We use the magnetic field measurements from the Fluxgate Magnetometer (FGM) \cite{russell_magnetospheric_2016}, the electric field measurements from the Electric field Double Probe (EDP) \cite{ergun_axial_2016,lindqvist_spin-plane_2016}, and the ion and electron distributions and their moments from the Fast Plasma Investigation (FPI) \cite{pollock_fast_2016}, the Hot Plasma Composition Analyser (HPCA) \cite{young_hot_2016}, the Fly Eye Energetic Particle Spectrometer (FEEPS) \cite{blake_flys_2016} and the Energetic Ion Spectrometer (EIS) \cite{mauk_energetic_2016}. The FPI instrument measures the ion and electron distributions in the thermal energy range $K\in [0.01-30~\textrm{keV}]$ assuming that the ion flux is fully dominated by the protons. We note that due to penetrating radiation effects, we correct the FPI-DIS moments according to the procedure described in \citeA{gershman_systematic_2019}. The HPCA instrument measures the ion flux over a broader energy range $K/q\in [0.01-40~\textrm{keV}/q]$ with a mass resolution $m/q\in [1-16~\textrm{a.m.u}/e]$ (i.e., H$^+$, He$^{+}$, He$^{2+}$ and O$^{+}$) using a carbon-foil based time-of-flight analyser. The FEEPS instrument measures ions, with no mass resolution, and electron flux in the energy ranges $K_i\in[40-1000~\textrm{keV}]$ and $K_e\in[20-1000~\textrm{keV}]$, respectively. The EIS instrument measures ion flux in the energy range $K/q\in [20-1000~\textrm{keV}]$ with time-of-flight based mass resolution (H$^+$, He$^{n+}$, and O$^{n+}$). For all instruments, we used both Fast Survey and Burst data.

We investigate a series of bursty bulk flows (BBFs) observed by the Magnetospheric Multiscale (MMS) spacecraft on July 23rd, 2017 between 16:10:00 UT and 18:10:00 UT. For this time period, the spacecraft were located in the magnetotail at (-24, 7, 4)~$R_E$ in Geocentric Solar Magnetospheric (GSM) coordinates, where $R_E$ is Earth's radius, as shown in Figure~\ref{fig:overview}a and \ref{fig:overview}b. The black lines in Figure~\ref{fig:overview}b show Earth's magnetic field lines computed using the T89 model \cite{tsyganenko_magnetospheric_1989}. During the time interval of the observations, the substorm activity was moderate with the AE index reaching $\sim 400$ nT.

During the event, the spacecraft separation shown in Figure~\ref{fig:overview}c, was $\Delta R \sim 10~\textrm{km} \sim 0.01 d_i$, where the ion inertial length $d_i~=~740~\textrm{km}$ is the typical thickness of a DF~\cite{runov_themis_2009}. As we are interested in energetic ions, which have the gyroradius much larger than the spacecraft separation scale, the fields measurements from FGM and EDP, and the ion moments from FPI-DIS are computed at the center of mass of the tetrahedron formed by the MMS spacecraft. Taking advantage of the small spacecraft separation, the FEEPS instruments onboard all spacecraft are combined together to provide a larger instantaneous field of view and better counting statistics. The same method was applied to EIS. Because of small discrepancies between the spacecraft, data from HPCA were taken from the MMS 2 spacecraft which offers the best agreement with other instruments.

\subsection{Overview}\label{ssec:overview}
\begin{figure}
    \centering
    \includegraphics[width=.969\linewidth]{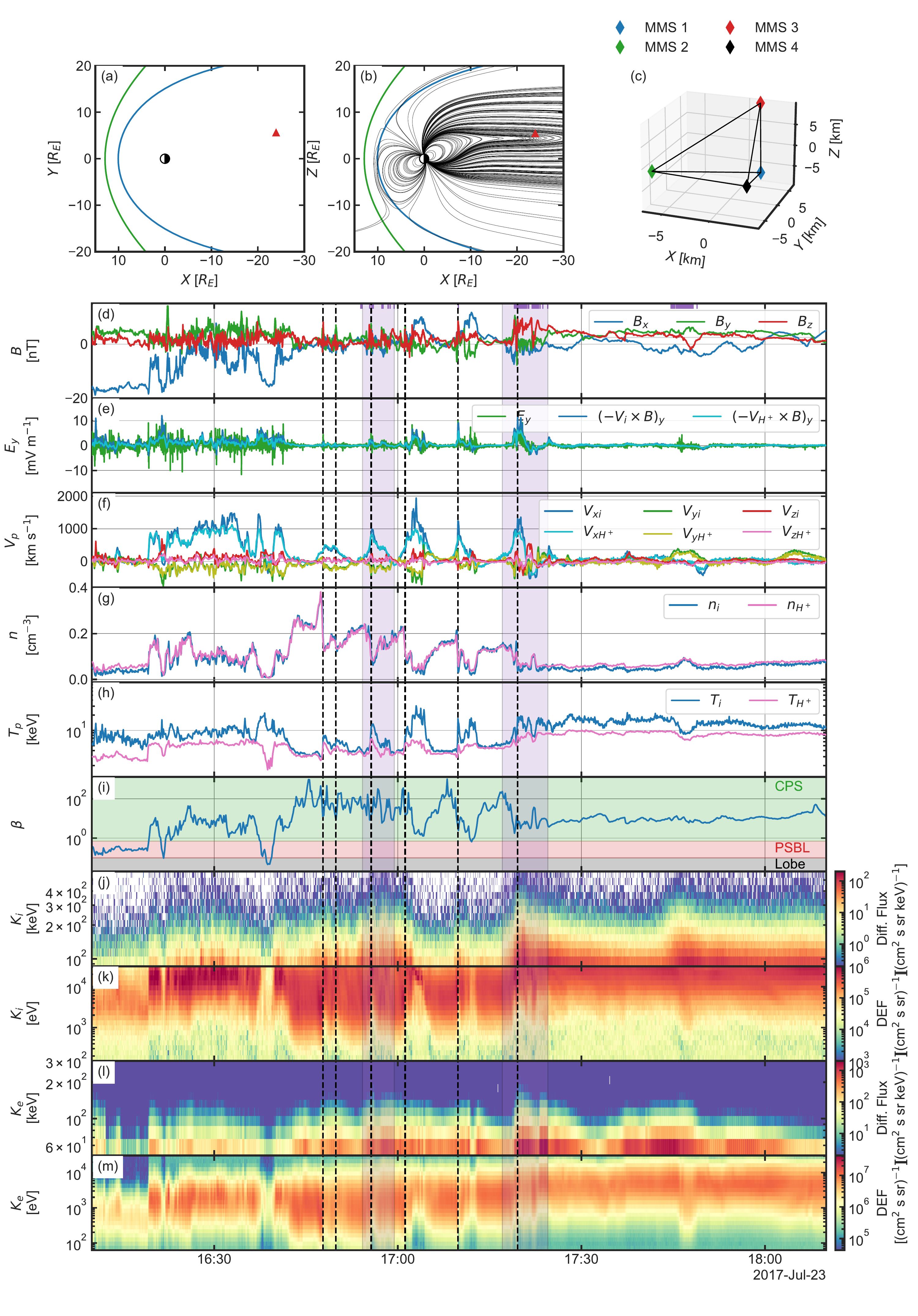}
    \caption{Overview of the event. (a)-(b) Location of the MMS spacecraft in GSE coordinates with the earth's bow shock (green), the earth's magnetopause (blue) and the magnetic field lines computed using the T89 model (black). (c) MMS tetrahedron configuration. (d) Magnetic field in GSM coordinates. (e) Dawn-dusk (GSM) electric field. (f) Proton bulk velocity in GSM coordinates. (g) Proton (blue and cyan) and alpha particles (green) number density. (h) Proton temperature. (i) Plasma parameter $\beta$. (j) Omni-directional ion differential particle flux. (k) Omni-directional ion differential energy flux. (l) Omni-directional electron differential particle flux. (m) Omni-directional electron differential energy flux. The purple dots in panel (d) are the energization times defined in Equation~\ref{eq:condition-flux}. The black dashed lines indicate the peaks in the energetic ion flux (panel (j)).}
    \label{fig:overview}
\end{figure}

An overview of the event is shown in Figure~\ref{fig:overview}. Initially, the magnetic field (Figure~\ref{fig:overview}d) is predominantly tailward (large negative $B_x$) which indicates that the MMS spacecraft were located southward of the magnetotail neutral sheet. Using the plasma moments from FPI, we plot in Figure~\ref{fig:overview}i, the plasma beta $\beta = p/p_{mag}$ with $p$ the plasma pressure and $p_{mag} = |\mathbf{B}|^2 / 2\mu_0$ the magnetic pressure. Using the criteria by \citeA{haaland_spectral_2010} to classify the regions, we observe that, consistent with the T89 model shown in Figure~\ref{fig:overview}b, the MMS spacecraft were initially in the plasma sheet boundary layer (PSBL, $0.01<\beta<0.2$) and then went into the central plasma sheet (CPS, $\beta>0.7$). We note two excursions into the lobe ($\beta < 0.01$) at 16:25:00 UT and 16:40:00 UT, suggesting a global north-south motion of the plasma sheet. Furthermore, we see that the Earthward magnetic field $B_x$ observed in the lobe at 16:40:00 UT is weaker than the one initially observed in the PSBL $\sim 20$ nT at 16:10:00 UT, which suggests a possible thinning/stretching of the plasma sheet possibly related to the growth phase of the substorm.

The MMS spacecraft observed a quasi-steady Earthward jet with $V_x \sim 1000~\textrm{km}~\textrm{s}^{-1}$ lasting until 16:40:00 UT, which is  followed by 5 Earthward BBFs (Figure~\ref{fig:overview}f). In particular, we note that the last BBF at 17:20:00 UT, exhibits oscillations of the Earthward ion bulk velocity with a period of $\sim 2~\textrm{min}$. Similar observations of BBF oscillations have been reported by \citeA{panov_multiple_2010} using the THEMIS spacecraft and by \citeA{merkin_contribution_2019} using the Lyon-Fedder-Mobarry (LFM) model run. \citeA{merkin_contribution_2019} concluded that these oscillations are a signature of a BBF overshooting its equilibrium and oscillating until settling in~\cite{chen_theory_1999}. At the leading edge of the BBFs, we observe in the FEEPS omni-directional ion and electron energy spectra plotted in Figures \ref{fig:overview}k and \ref{fig:overview}m respectively, strong sharp energization of the thermal ions and electrons. Associated with the energization of the thermal particles, we observe a flux increase of the supra-thermal ions and electrons in Figures \ref{fig:overview}j and \ref{fig:overview}l respectively. We note that energetic ions and electrons are also seen outside of the BBFs but since the BBFs are the primary subject of the paper, we focus on those observed at the leading edge of the Earthward flows. We marked the BBF-associated peaks in the energetic ion flux by the dashed black lines. The peaks are associated with large drops in the ion density (Figure~\ref{fig:overview}g), increases in the ion temperature (Figure~\ref{fig:overview}h), and  dipolarizations (increased $B_z$) of the magnetic field (Figure~\ref{fig:overview}d), which are followed by large amplitude electric field activity (Figure~\ref{fig:overview}e). Based on these properties we identify these structures as DFs.

In order to focus the study of ion energization processes, we select intervals based on the supra-thermal ion flux at 200 keV. Using the FEEPS omni-directional ion energy spectrum, which does not separate between the different ion masses, we identify the energization times as:

\begin{equation}
    J^{200} > 2 \sigma_{J^{200}} = 2\sqrt{\frac{\sum_{i=1}^n \left ( J_i^{200} - \bar{J}^{200}\right )}{N - 1}}
    \label{eq:condition-flux}
\end{equation}
where $J^{200}$ is the ion flux at $200~\textrm{keV}$, $\bar{J}^{200}$ its mean value over the 2 hours interval 16:10:00 - 18:10:00 UT  and $N$ the number of time records. The condition defined by Equation~\ref{eq:condition-flux} corresponds statistically to a 95th percentile threshold in a normal distribution. The marked times at the top of Figure~\ref{fig:overview}d, reveals three clusters with two that are associated with a high speed $V_x > 500~\textrm{km}~\textrm{s}^{-1}$ jet. Since the third cluster is not associated with a BBF, we focus only on the two first intervals henceforth refered to as event I (16:54:14-16:59:26 UT) and event II (17:17:04-17:22:36 UT). These two intervals are indicated by purple shading in Figures~\ref{fig:overview}d to \ref{fig:overview}m. In the rest of the paper, we focus on the ion acceleration during these two intervals.

\subsection{Ion Composition and Charge-State-Dependent Acceleration} \label{ssec:ion-composition}
\begin{figure}
    \centering
    \includegraphics[width=\linewidth]{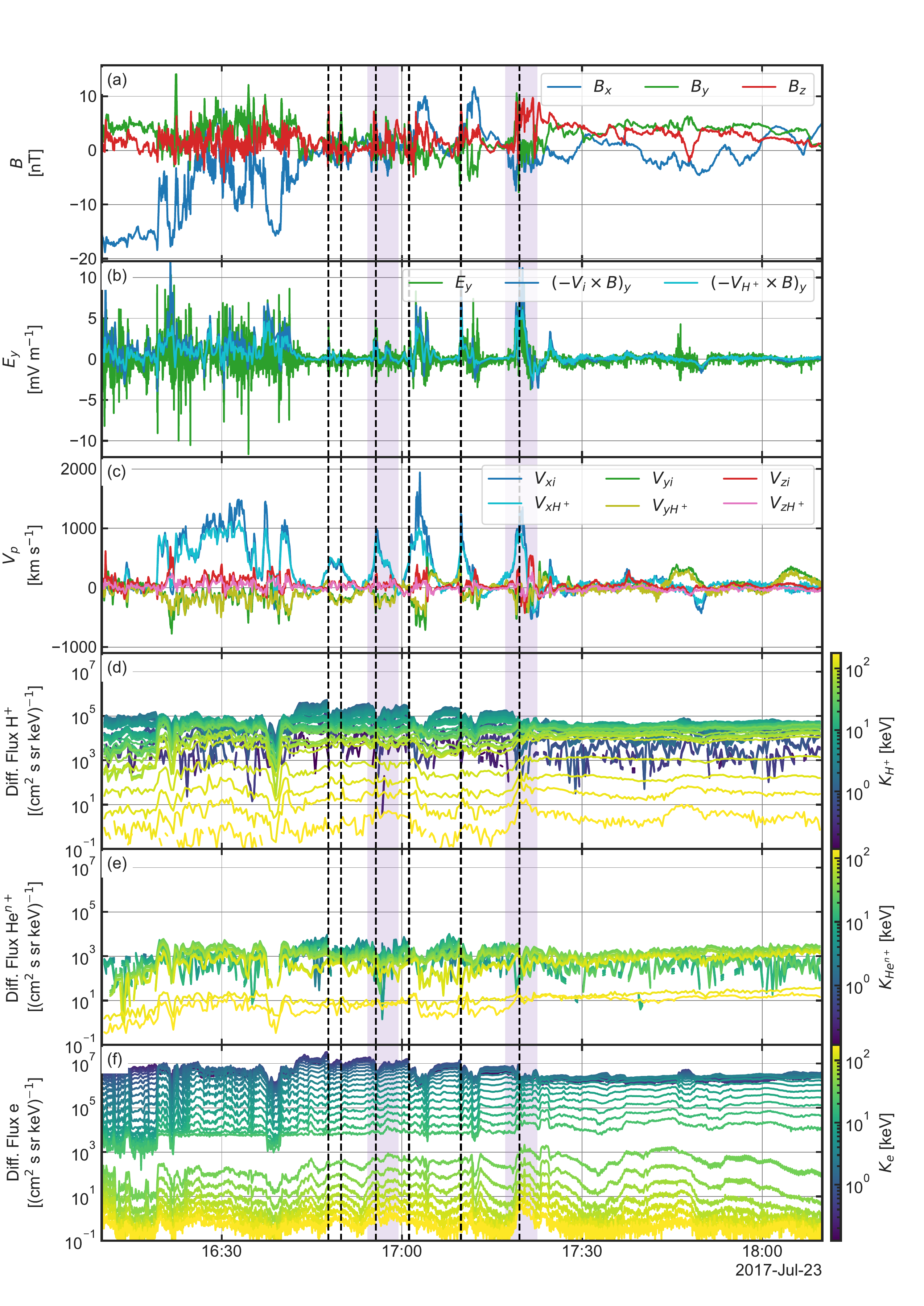}
    \caption{Ion (H$^{+}$ and He$^{n+}$) and electron flux changes. (a) Magnetic field in GSM coordinates. (b) Dawn-dusk cross-tail electric field from EDP (green), FPI-DIS $\mathbf{V_{i}}\times \mathbf{B}$ (blue) and HPCA $\mathbf{V_{H^{+}}}\times \mathbf{B}$ (cyan). (c) Ion bulk velocity from FPI-DIS and HPCA. (d) H$^+$ flux time series from combined FPI-DIS and EIS. (e) He$^{n+}$ flux time series from combined HPCA and EIS. (f) Electron flux time series from combined FPI-DES and EIS. The black dashed lines indicate the injections.}
    \label{fig:ion-flux-tseries}
\end{figure}

In order to characterize the ion acceleration, it is important to know the composition of the supra-thermal ions. We plot in Figure~\ref{fig:ion-flux-tseries}d the time series of the Fast Survey Mode proton flux at different energies. At low (thermal) energies $K_p < 40~\textrm{keV}$, we use the proton H$^+$ flux from the HPCA instrument. At higher energies $K_p > 20~\textrm{keV}$ we use the proton H$^+$ flux from the EIS instrument. We also plot in Figure~\ref{fig:ion-flux-tseries}e the Fast Survey Mode Helium flux. At low energies ($K_{He^{n+}} < 80~\textrm{keV}$) we use the He$^{2+}$ flux from HPCA, which offers both mass and charge state resolution. At supra-thermal energies, we use the He$^{n+}$ flux from the EIS instrument which, unlike HPCA, does not provide charge state resolution. We note that, although HPCA also provides the He$^+$ flux, for this event no significant He$^+$ flux was observed at energies $< 40~\textrm{keV}$. The O$^{n+}$ ions also measured by the EIS instrument are not treated in this study due to low statistics (i.e. large errors $\delta f/f = 1 / \sqrt{n}$, with $n$ the number of counts assuming that the counting statistics are described by Poisson's statistics) in the EIS energy range. Finally, we plot in Figure~\ref{fig:ion-flux-tseries}f the Fast Survey Mode electron flux, where the low energies ($K_e < 30~\textrm{keV}$) are measured by the FPI-DES instrument while energies $K_e > 30~\textrm{keV}$ are measured by the FEEPS instrument.

At energies $K < 10~\textrm{keV}$, the electron flux (Figure~\ref{fig:ion-flux-tseries}d) dominates the ion (proton and heavy ions) flux (Figure~\ref{fig:ion-flux-tseries}e). Similar intensity dominance of $< 10~\textrm{keV}$ electrons, have been observed by \citeA{runov_average_2015} using the THEMIS spacecraft. At energies $10~\textrm{keV} < K < 150~\textrm{keV}$, the flux is dominated by the proton flux, while at higher energies $K > 150~\textrm{keV}$ the flux is dominated by the helium flux. This result is consistent with observation at $6 < L < 16.5$ by \citeA{cohen_dominance_2017}, who showed the intensity dominance for $K > 150~\textrm{keV}$ of multiply-charged heavy ions in the magnetosphere.

In order to identify the charge state of the observed He$^{n+}$ ions from EIS measurement, we use the correlation technique introduced by \citeA{mitchell_energetic_2018} for Van Allen Probes RBSPICE measurements, and later employed for MMS EIS measurements by \citeA{bingham_chargestatedependent_2020,bingham_evidence_2021}. From the bounce-averaged guiding center description of particle motion \cite{kistler_energy_1989}, 

\begin{equation}
    \frac{\textrm{d}\mathbf{x}}{\textrm{dt}} = \mathbf{v_{E\times B}} + \frac{K_{\perp}}{q B}\frac{\mathbf{B}\times \nabla\mathbf{B}}{\mathbf{B}^2}+ \frac{2K_{\parallel}}{q B} \frac{\mathbf{R_c}\times \mathbf{B}}{\mathbf{R_c}^2 B}
    \label{eq:guiding-center}
\end{equation}
where $\mathbf{x}$ is the position of the guiding center, $\mathbf{v_{E\times B}} = \mathbf{E}\times \mathbf{B} / |\mathbf{B}^2|$ is the drift velocity and $\mathbf{R_c}$ is the radius of curvature of the magnetic field, it follows that the ions with the same $K/q$ follow the same trajectories and are energized in a charge-state-dependant manner. Due to the charge state dependence of the second and third terms of Equation~\ref{eq:guiding-center}, \citeA{schulz_particle_1974} showed that the temporal flux changes are ordered by $K/q$. Using this result, \citeA{mitchell_energetic_2018}, used the correlation of the temporal changes of the Van Allen Probes RBSPICE measurements of the proton and heavy ion fluxes to infer the charge state of the heavy ions. A large cross-correlation coefficient between the proton flux at the energy $K_{H^+}$ and the flux of the specie $s$ at the energy $K_{s} = q_s K_{H^+}$, would suggest that the most likely charge state of the species $s$ is $q_s$.

\begin{figure}
    \centering
    \includegraphics[width=\linewidth]{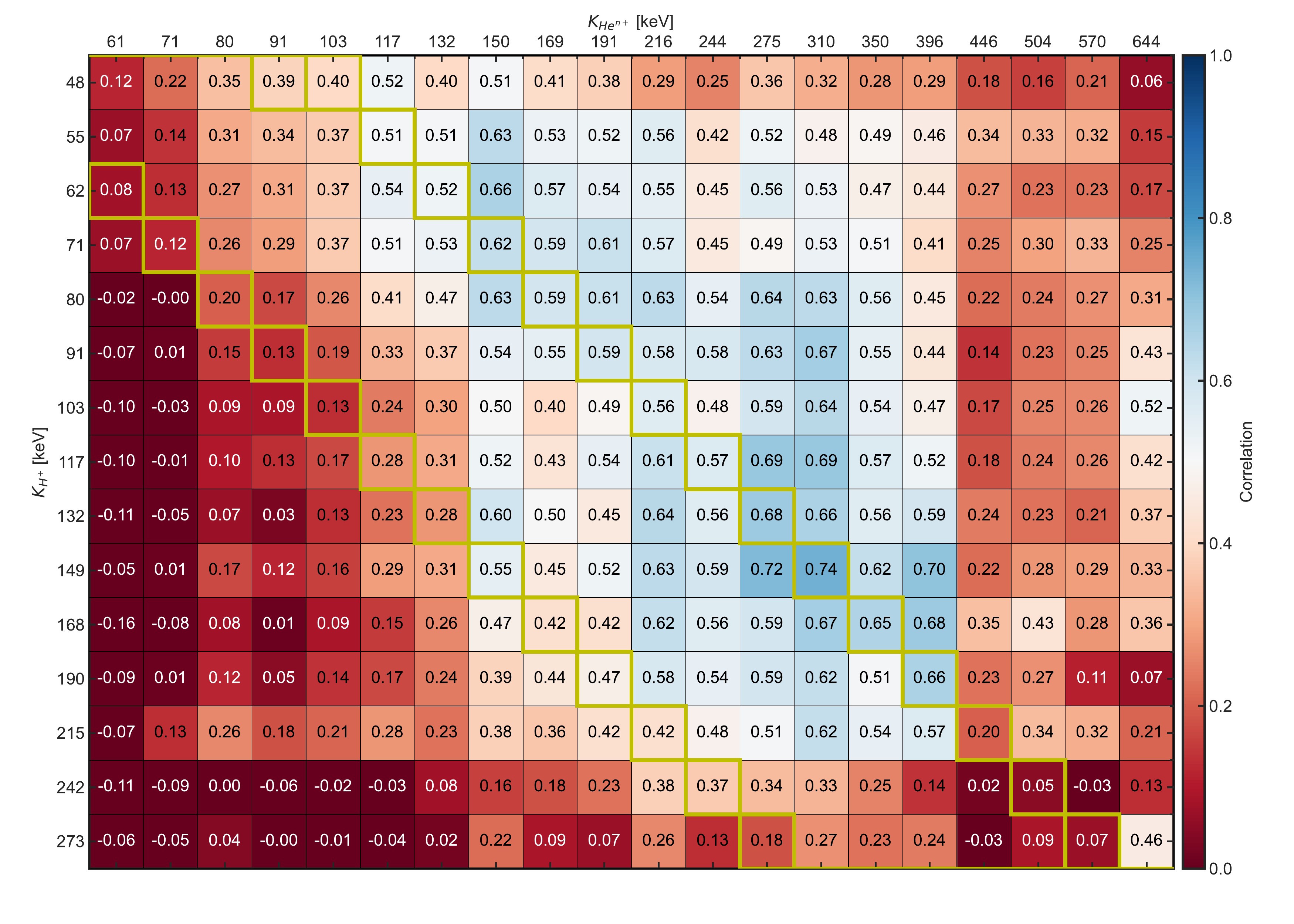}
    \caption{Cross-correlation coefficients between the time series of H$^+$ and He$^{n+}$ from EIS measurements at different energies taken between 16:10-18:10 UT on July 23, 2017. The yellow boxes indicate $K_{He^{n+}} = 1*K_{H^{+}}$ (lower left diagonal) and $K_{He^{n+}} = 2*K_{H^{+}}$ (upper right diagonal)}
    \label{fig:ion-correlation}
\end{figure}

On large scales, we observe that the flux changes of the protons (Figure~\ref{fig:ion-flux-tseries}d) and the Helium ions (Figure~\ref{fig:ion-flux-tseries}e) are qualitatively well correlated. We note that every injection (supra-thermal ions flux increase at the DFs) marked by the black dashed lines is dispersionless (i.e., the flux increase occurs at the same time at all energies). In Figure~\ref{fig:ion-correlation} we plot the cross-correlation coefficient between the changes in the Burst mode H$^+$ and He$^{n+}$ fluxes during the time interval 16:10:00 - 18:10:00 UT. The total length of the Burst data is about 1 hour ($\sim 180$ omni-directional spin-averaged points) with some gaps. In order to guide the eye, we mark with yellow boxes the energies that satisfy the relation $K_{He^{n+}} = q*K_{H^{+}}$ where $q = 1$ corresponds to He$^+$ and $q=2$ to He$^{2+}$. We observe a ridge of high correlation which falls within the yellow boxes corresponding to $q = 2$, indicating that on large scales the dynamics is ordered by $K/q$, and that the dominant charge state of the helium is likely He$^{2+}$, consistent with the aforementioned observation of He$^{2+}$ and no He$^+$ at thermal energies ($K/q < 40~\textrm{keV}/q$). This result indicates that during our event in the magnetotail, the heavy ions are of solar wind origin, consistent with the MMS observations by \citeA{bingham_chargestatedependent_2020,bingham_evidence_2021}.

To investigate whether the dynamics are ordered by $K/q$ on local scales, we compare the flux enhancement for H$^+$ and He$^{n+}$ as a function of the energy per charge $K/q$ for the two selected intervals. To do so, we assume that the source population is upstream of the DF. Indeed, as shown in the combined MHD/test particle simulations by \citeA{birn_particle_2013,birn_energetic_2015}, the main source of accelerated ions is the central plasma sheet. Furthermore, we can assume that the plasma sheet is rather homogeneous in the dawn-dusk direction on the scale of the flow channel ($1-5~R_E$). Hence, we assume that the source population is the plasma sheet at rest far ahead of the acceleration site (i.e., the DF). We emphasize the importance of using a source region far from the DF, at least a few ion gyroradii, because of possible contamination by the DF-reflected ions ahead of the DF \cite{zhou_accelerated_2010,zhou_nature_2011}. On the other hand, the energized distribution is taken at the time of peak of the proton flux at $K_p \approx 70~\textrm{keV}$. 

\begin{figure}
    \centering
    \includegraphics[width=\linewidth]{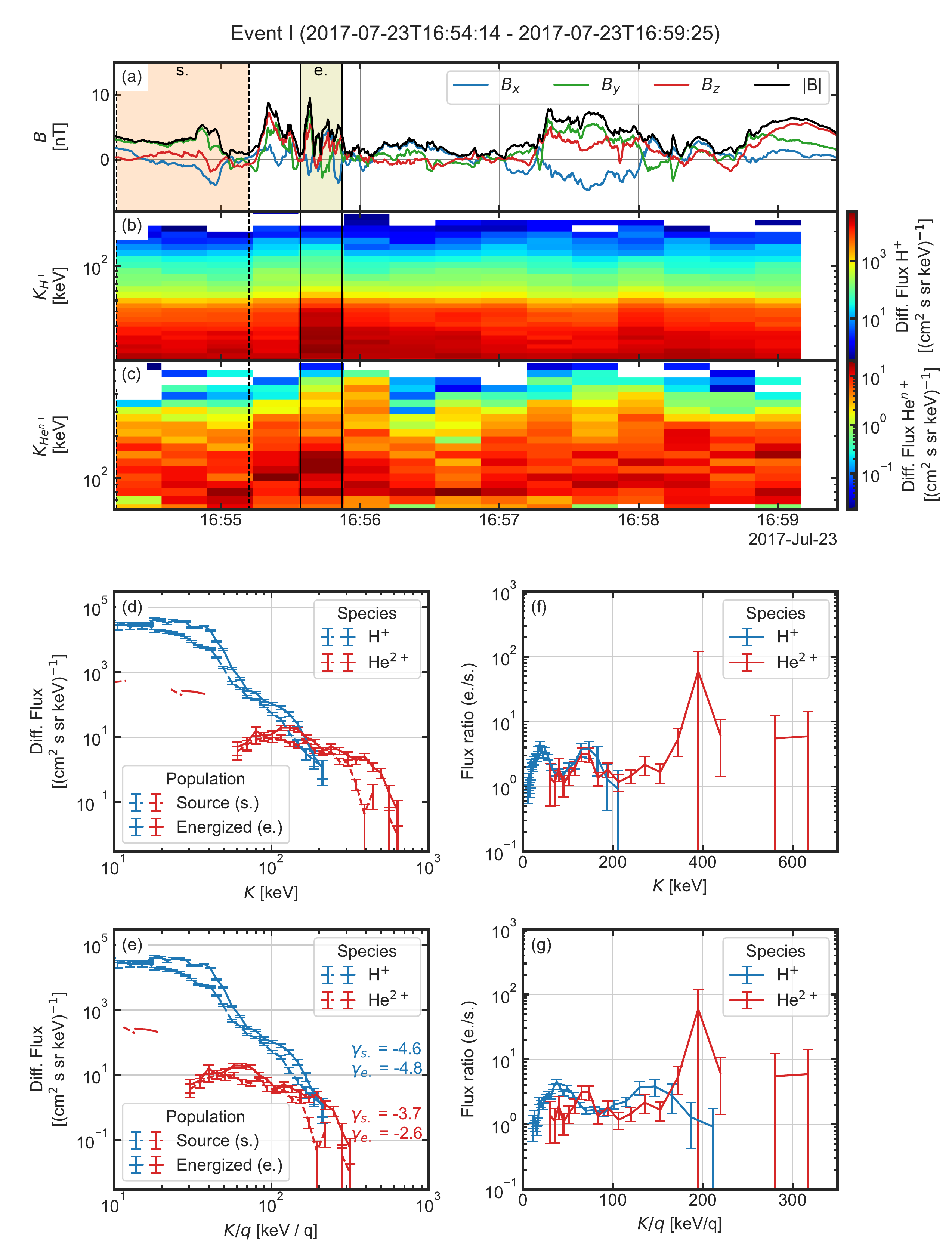}
    \caption{H$^{+}$ and He$^{n+}$ source and energized flux spectra for event I. (a) Magnetic field in GSM coordinates. (b) EIS H$^+$ omni-directional flux energy spectrum. (c) EIS He$^{n+}$ omni-directional flux energy spectrum. H$^{+}$ (blue) and He$^{n+}$ (red) flux energy spectra in the source (dashed line) and energized (solid line) regions as a function of the energy $K$ (d) and the energy per charge $K/q$ (e). H$^{+}$ (blue) and He$^{n+}$ (red) energized to source flux ratio energy spectra as a function of the energy $K$ (f) and the energy per charge $K/q$ (g).}
    \label{fig:ion-spectra-1}
\end{figure}

For event I (16:54:14-16:59:26 UT), we observe that the magnetic field is highly variable with a chain of small scale structures in Figure~\ref{fig:ion-spectra-1}a. The characteristic time scale of the magnetic field structures is $\tau = 5$ s. Using the average ion bulk velocity associated with the magnetic field structures $V_i = 830 \pm 110~\textrm{km}~\textrm{s}^{-1}$, a characteristic spatial scale of $L_x = 0.65\pm 0.08~R_E$ is calculated in the Earth-tail direction. We plot the EIS proton (blue) and combined HPCA/EIS helium (red) omni-directional energy spectra in Figure~\ref{fig:ion-spectra-1}b and \ref{fig:ion-spectra-1}c respectively. The time interval where we take the source and energized ion distributions are marked as "s." and "e.". We plot in Figure~\ref{fig:ion-spectra-1}d and \ref{fig:ion-spectra-1}e the source (dashed lines) and energized (solid lines) enhancement omni-directional flux as a function of the ion energy $K$ and ion energy per charge $K/q$, respectively, and the ratio of the energized to source fluxes in Figures~\ref{fig:ion-spectra-1}f and \ref{fig:ion-spectra-1}g. As already mentioned, at energies $ > 150~\textrm{keV}$ the helium flux dominates the proton flux (Figure~\ref{fig:ion-spectra-1}d). 

The spectral slope of the flux as a function of the energy per charge plotted in Figure~\ref{fig:ion-spectra-1}e, differs significantly between the species. In particular, at energies $> 50~\textrm{KeV}/q$, the Helium (He$^{2+}$) spectrum is harder (i.e., has a smaller spectral slope $|\gamma_s| = 3.7,~|\gamma_e| = 2.6$), than that of the protons ($|\gamma_s| = 4.6,~|\gamma_e| = 4.8$). To compute the energy per charge ratio, we assumed that the helium is in the alpha charge state He$^{2+}$, based on the above charge state analysis. When plotted as a function of the energy $K$, we observe that the flux ratio of the He$^{2+}$ agrees with the one of the protons (Figure~\ref{fig:ion-spectra-1}f). On the other hand, when plotted as a function of the energy per charge $K/q$, we observe that the flux ratio of the He$^{2+}$ ions does not agree with the one of the protons (Figure~\ref{fig:ion-spectra-1}g). This results indicates that at constant $K/q$, the flux changes of the He$^{2+}$ ions differs from those of the protons, and thus the flux changes are not ordered by $K/q$, meaning that the energization does not depend solely on the ion charge state.

\begin{figure}
    \centering
    \includegraphics[width=\linewidth]{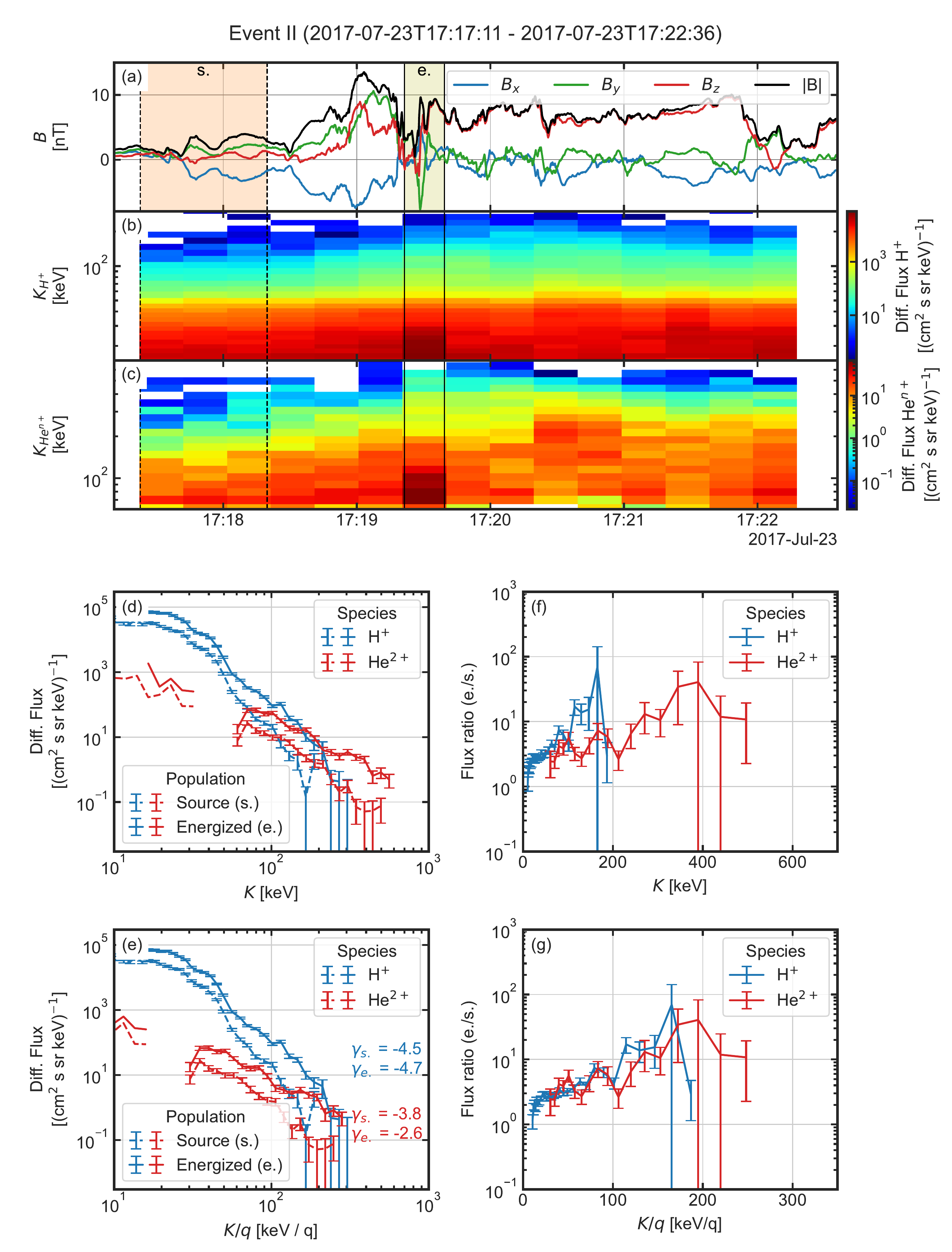}
    \caption{H$^{+}$ and He$^{n+}$ source and energized flux spectra for event II. (a) Magnetic field in GSM coordinates. (b) EIS H$^+$ omni-directional flux energy spectrum. (c) EIS He$^{n+}$ omni-directional flux energy spectrum. H$^{+}$ (blue) and He$^{n+}$ (red) flux energy spectra in the source (dashed line) and energized (solid line) regions as a function of the energy $K$ (d) and the energy per charge $K/q$ (e). H$^{+}$ (blue) and He$^{n+}$ (red) energized to source flux ratio energy spectra as a function of the energy $K$ (f) and the energy per charge $K/q$ (g).}
    \label{fig:ion-spectra-2}
\end{figure}

For event II (17:17:04-17:22:36 UT), the magnetic field (Figure~\ref{fig:ion-spectra-2}a) in the wake of the DF has a stable configuration suggesting a large-scale structure with a spatial scale of $\sim 6~R_E$ in the Earth-tail direction. The spectral slopes of the H$^+$ and He$^{2+}$ ions are similar for both source and energized ions (Figure~\ref{fig:ion-spectra-2}e). Such alignment of the spectral shape of different species have been observed using MMS by \citeA{bingham_chargestatedependent_2020}. Moreover, the flux ratio of the two species shows a very good agreement when plotted as a function of $K/q$ in Figure~\ref{fig:ion-spectra-2}g. This indicates that the energization likely depends solely on the charge state, which suggests direct acceleration of ions by the inductive electric field which is mass independent.

\subsection{Acceleration of Protons}\label{ssec:acc-mechanism}
\begin{figure}
    \centering
    \includegraphics[width=\linewidth]{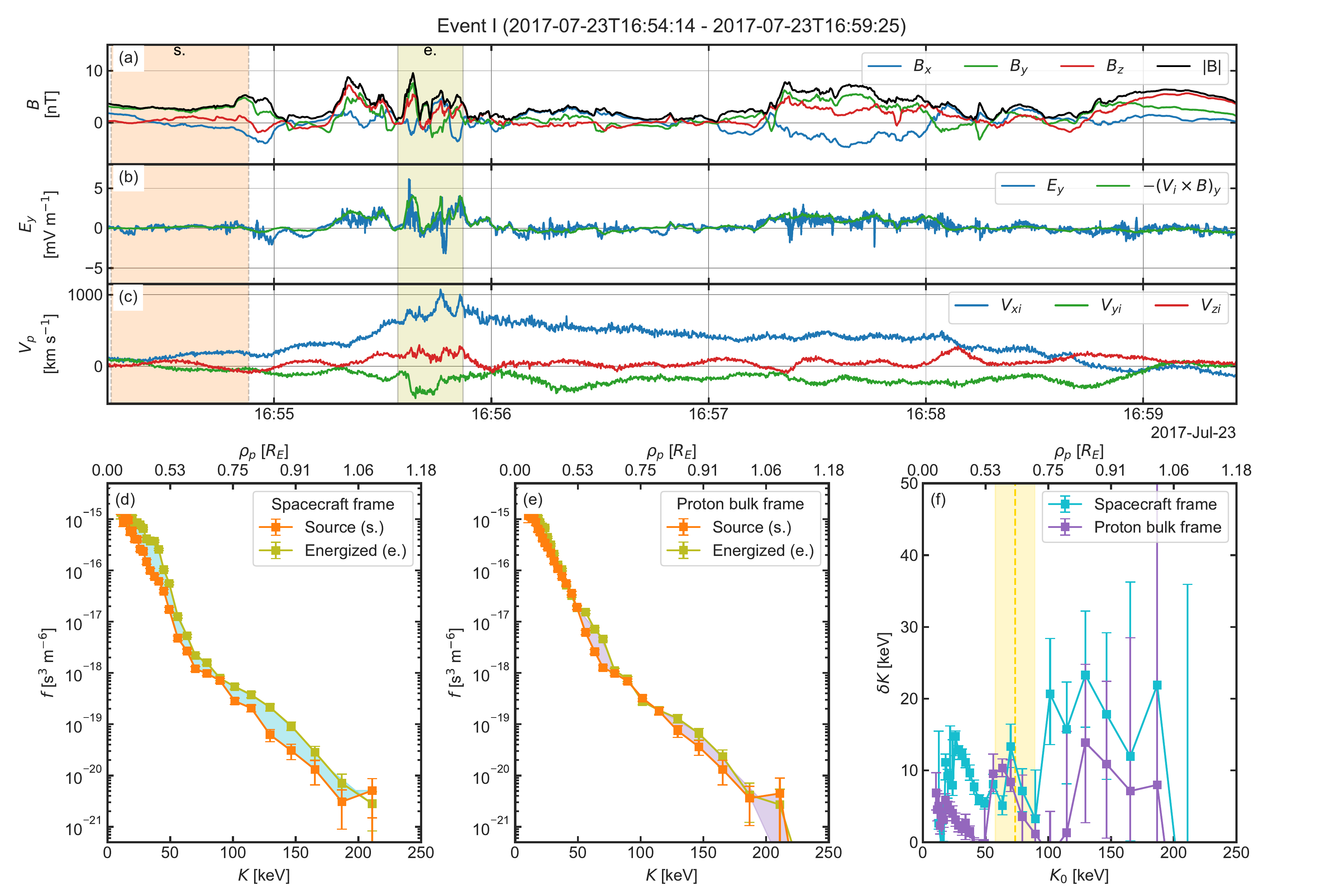}
    \caption{Proton flux energy spectrum in the spacecraft and proton frame for event I. (a) Magnetic field in GSM coordinates. (b) Dawn-dusk cross-tail electric field from EDP (blue) and $-V_i\times B$ (green). (c) Ion bulk velocity in GSM coordinates. The orange and green shaded regions in panels (a)-(c) show the source and energized regions. (d) Source (orange) and energized (green) proton phase space density energy spectrum in the spacecraft frame. The cyan area emphasize the difference between the two lines. (e) Source (orange) and energized (green) proton phase space density energy spectrum in the proton bulk frame. The purple area emphasize the difference between the two lines. (f) Energy increase in the spacecraft frame (cyan) and proton bulk frame (purple). The gold line shows the energy of a proton with a gyroradius of $\rho_p = L_{pulse}$ (see text).}
    \label{fig:acc-mechanism-1}
\end{figure}

We now investigate the acceleration mechanisms responsible for the energetic ion enhancement for both intervals. Here, we focus on the proton acceleration mechanism since as shown in Figures~\ref{fig:ion-spectra-1} and \ref{fig:ion-spectra-2}, the larger statistics i.e., smaller errors are measured for protons. We plot in Figure~\ref{fig:acc-mechanism-1}d the source (orange) and energized (green) omni-directional proton phase space density energy spectra in the spacecraft frame. The phase space density $f$ is computed from the proton flux measured by EIS using the standard relation between differential flux and phase space density (non-relativistic case)

\begin{equation}
    \label{eq:flux2psd}
    f = \frac{m_p^2}{2K}J,
\end{equation}

\noindent where $m_p$ is the proton mass and $K$ is the effective energy of the EIS energy channel. We observe that the two spectra are nearly parallel with some minor discrepancies. 

To account for the effect of the bulk motion of the ion jet, we transform the proton velocity distribution into the proton bulk frame. To do so, we interpolate the proton distribution defined on the 3D spherical velocity space grid onto the same grid translated along the proton bulk velocity. Figure~\ref{fig:acc-mechanism-1}e shows the omni-directional phase space density energy spectra in the reference frame of the proton bulk flow. We observe good agreement between the source and energized distributions at energies $< 50~\textrm{keV}$, which suggests that the energization of the low energy protons is due to the bulk flow.

The energy gain $\delta K$ at the same PSD level highlighted by the cyan-shaded region in Figure~\ref{fig:acc-mechanism-1}d and the purple-shaded region in Figure~\ref{fig:acc-mechanism-1}e are plotted in Figure~\ref{fig:acc-mechanism-1}f (cyan and purple squares) as a function of the initial energy $K_0$. We observe that the energy gain $\delta K$ oscillates as a function of the initial energy $K_0$ ($\sigma_{\delta K} / \langle \delta K \rangle= 0.91$). In particular, at energies $< 50~\textrm{keV}$, $\delta K = 6.8\pm 4.8~\textrm{keV}$ which corresponds to $V_i=1139\pm 407~\textrm{km}~\textrm{s}^{-1}$, consistent (within $0.74\sigma$) with the average proton bulk speed $V_i = 830\pm 110~\textrm{km}~\textrm{s}^{-1}$ observed in Figure~\ref{fig:acc-mechanism-1}c.

In the proton bulk frame for energies $> 50~\textrm{keV}$ we observe a significant energy gain in Figure~\ref{fig:acc-mechanism-1}f. In particular, we observe a peak of the energy gain $\delta K$ in the proton bulk frame (purple squares) at $K_0 \sim 140~\textrm{keV}$ in Figure~\ref{fig:acc-mechanism-1}f. However, the large errors in the proton phase-space densities $f$ at energies $K_0 > 140~\textrm{keV}$ yield large uncertainties on the energy gain $\delta K$, so that it is difficult to conclude if the energy gain at $K_0 \sim 140~\textrm{keV}$ is a real physical enhancement. On the other hand, we observe a bulge in the energy gain at $K_0 = 66.6\pm 7.3~\textrm{keV}$ in Figure~\ref{fig:acc-mechanism-1}f. We estimate the characteristic time scale $\tau = 5~\textrm{s}\approx 0.68~f_{cp}^{-1}$ of the magnetic field structures (Figure~\ref{fig:acc-mechanism-1}b). Using the average ion bulk velocity associated with the magnetic field structure, it yields a spatial scale of $L_{x} = 0.65\pm 0.08~R_E$. The energy of a proton with a gyroradius so that $\rho_p = L_{x}$ is $K_p = 74.7 \pm 19.7~\textrm{keV}$ (gold line), consistent with the energy $K_0 = 66.6\pm 7.3~\textrm{keV}$ of the bulge of energy gain.

\begin{figure}
    \centering
    \includegraphics[width=\linewidth]{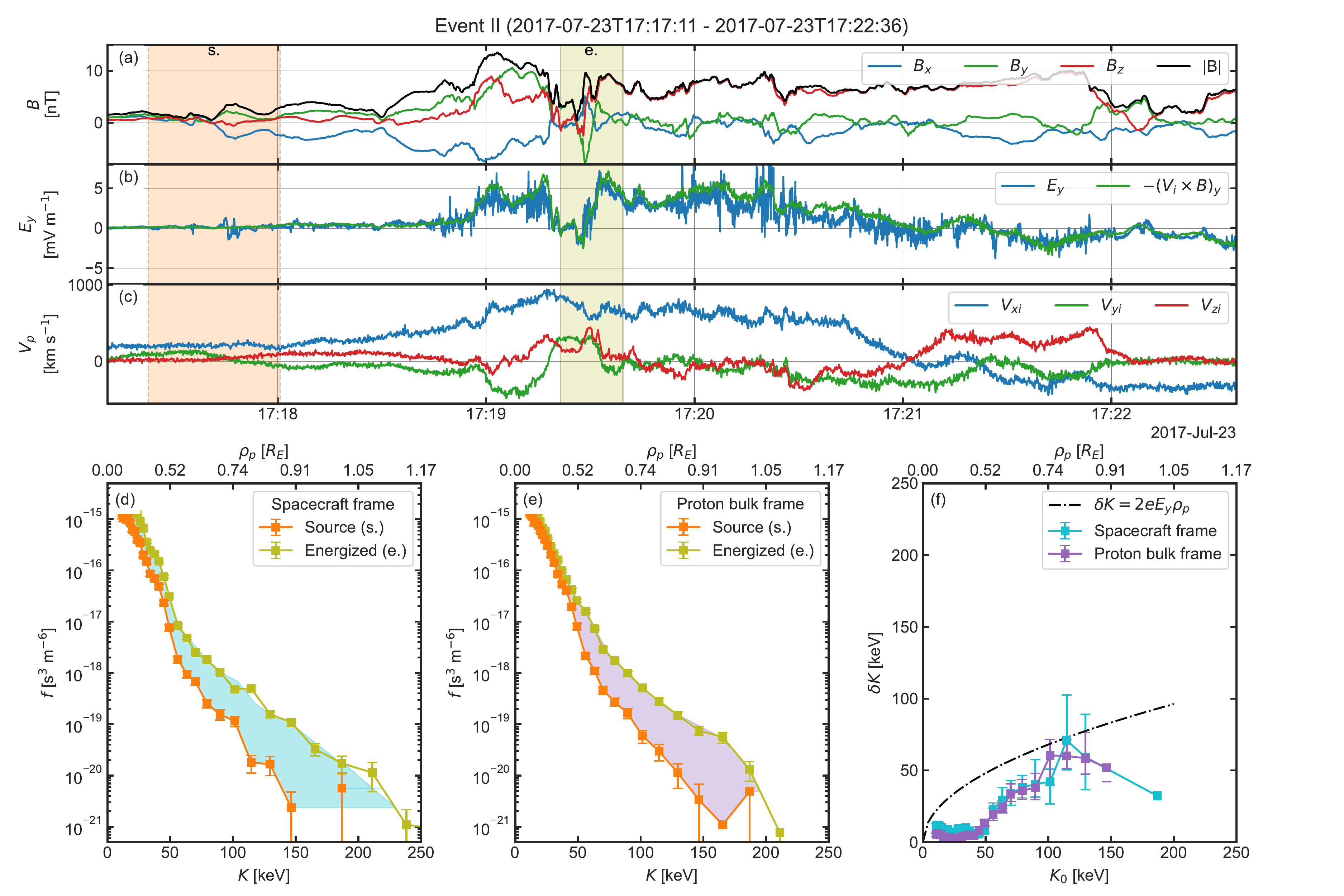}
    \caption{Proton flux energy spectrum in the spacecraft and proton frame for event II. (a) Magnetic field in GSM coordinates. (b) Dawn-dusk cross-tail electric field from EDP (blue) and $-V_i\times B$ (green). (c) Ion bulk velocity in GSM coordinates. The orange and green shaded regions in panels (a)-(c) show the source and energized regions. (d) Source (orange) and energized (green) proton phase space density energy spectrum in the spacecraft frame. The cyan area emphasize the difference between the two lines. (e) Source (orange) and energized (green) proton phase space density energy spectrum in the proton bulk frame. The purple area emphasize the difference between the two lines. (f) Energy increase in the spacecraft frame (cyan) and proton bulk frame (purple). The black dash-dotted line shows the model $\delta K = 2eE_y\rho_p$. }
    \label{fig:acc-mechanism-2}
\end{figure}

Figure~\ref{fig:acc-mechanism-2} shows data for the event II (17:17:04-17:22:36 UT) in the same format as Figure~\ref{fig:acc-mechanism-1}. We observe in Figure~\ref{fig:acc-mechanism-2}d that at energies $< 50~\textrm{keV}$, the energy gain $\delta K$ in the spacecraft frame is rather constant and does not depend on the initial energy $K_0$. This result is also seen in Figure~\ref{fig:acc-mechanism-2}f, where we observe a constant energization $\delta K = 8.4\pm1.9~\textrm{keV}$ (cyan squares) at energies $K_0 < 50~\textrm{keV}$. This constant energization corresponds to $V_i=1268 \pm 143~\textrm{km}~\textrm{s}^{-1}$, which is consistent (within $1.95\sigma$) with the 20 s (17:19:20-17:19:40 UT) average of the ion bulk speed $V_i=732\pm235~\textrm{km}~\textrm{s}^{-1}$, and comparable to the maximum ion bulk speed $V_i^{max}=1001~\textrm{km}~\textrm{s}^{-1}$.

To account for the effect of the bulk flow, we transform the proton distribution into the proton bulk frame using the 20 s (17:19:20-17:19:40 UT) average bulk velocity and plot the omni-directional phase space density energy spectrum in Figure~\ref{fig:acc-mechanism-2}e. We observe good agreement between the source (orange) and the energized (green) spectra at low energies $< 50~\textrm{keV}$, and a negligible energy gain $\delta K$ in the proton bulk frame shown in Figure~\ref{fig:acc-mechanism-2}f. This result suggests that, similar to the event I (16:54:14-16:57:25 UT), the energization of the low energy proton $< 50~\textrm{keV}$ is due to the proton bulk flow. 

On the other hand, for protons with initial energy $> 50~\textrm{keV}$ (i.e., $\rho_p > 0.52~R_E$), we observe a significant energy gain in the proton bulk frame in Figures~\ref{fig:acc-mechanism-2}e and~\ref{fig:acc-mechanism-2}f, which suggests that protons with a gyroradius $\rho_p < 0.52~R_E$ have a zero net energy gain in the proton bulk frame, while protons with a gyroradius $\rho_p > 0.52~R_E$ gain energy. This net energy gain suggests that, for protons with a gyroradius $\rho_p < 0.52~R_E$ the energy gained from the cross-tail electric field during the duskward part of the gyration is compensated by the energy loss during the dawnward part of the gyration. On the other hand, for a proton with a gyroradius $\rho_p \gtrsim 0.52~R_E$ part of the decelerating dawnward part of the proton gyration is outside of the electric field region, and thus does not compensate the energization during the duskward part of the orbit. Preferential acceleration of protons with $\rho_p \gtrsim 0.52~R_E$ indicates that the electric field structure, which is the acceleration region, is spatially limited. In Figure~\ref{fig:acc-mechanism-2}f we plot (black dashed line) the model proposed by \citeA{artemyev_acceleration_2015}, of the acceleration of protons in a spatially limited electric field pulse $\delta K = 2eE_y\rho_p\propto \sqrt{K_0}$. We observe that at energies $K_0 > 100~\textrm{keV}$, the model agrees with the measured energy gain, but at energies $50~\textrm{keV} < K_0 < 100~\textrm{keV}$, because the proton orbit in the acceleration region is much more complicated than a simple half gyration described by \citeA{artemyev_acceleration_2015}, the model overestimates the energy gain.

\section{Discussion}
\label{sec:discussion}
We found that for both the small-scale turbulent and the large-scale magnetic field structures, at energies $< 50~\textrm{keV}$ the proton energization is of the order of the proton bulk energy $\delta K \sim K_{bulk}$ in the spacecraft frame, so that $\delta K \sim 0$ in the proton frame. This suggests that protons with energies $< 50~\textrm{keV}$ gyrate inside the $E \times B$ earthward convected magnetic field structure with no net energy gain. In this scenario, the energy gain due to the interaction with the cross-tail motional electric field during the first half (duskward) of the gyration, is suppressed during the second half (dawnward) where the particle is slowed down by the same electric field. So that, over one gyration no net energy is gained by the ion. Hence, we conclude that ions with a gyroradius $\rho_p < L_y$ with $L_y$ is the dawn-dusk scale of the structure, are $E \times B$ drifting without gaining energy from the interaction with the electric field.

In the event I (16:54:14-16:57:25 UT), we found that the ion flux changes associated with the acceleration are not ordered by $K/q$, which implies that the energization is non-adiabatic. Indeed, the ion motion is adiabatic when the scales (both temporal $\tau$ and spatial $L_x,~L_y$) of the electromagnetic fluctuations are larger than the ion scales (gyroperiod $f_{cp}^{-1}$ and gyroradius $\rho_p$). In such case Equation~\ref{eq:guiding-center} implies that the energization is ordered by $K/q$. For the event I, we observe that the flux changes are not ordered by $K/q$. Hence, the energization process does not depend solely on the charge state, which implies that the energization is non-adiabatic. 

The later result is also supported by our observation of electromagnetic fluctuations on temporal scales comparable to the ion gyroperiod, which contributes to the violation of the first adiabatic invariant. In the event I (16:54:14-16:57:25 UT), we observe that only the ions with gyroradius of $\rho_p \sim L_x$, where $L_x$ is the Earth-tail scale of the electromagnetic field structure, experience significant energization. This preferential energization suggests that the electromagnetic energy injected in the form of large scale structures (BBF) cascades down to the ion scales where it dissipates in the form of particle acceleration. Using Cluster,~\citeA{malykhin_contrasting_2018} observed a similar bulge in the proton energy spectrum in the $70~\textrm{keV}-90~\textrm{keV}$ energy range. They explained the formation of such bulge by the non-adiabatic resonant interaction of thermal protons with the DFs \cite{artemyev_ion_2012}. Using the analytical model of non-adiabatic resonant interaction proposed by \citeA{artemyev_ion_2012}, we can estimate the dawn-dusk scale of the DF as $L_y = W / |q E_y|$. We estimate the dawn-dusk scale of the DF to be $L_y = 1.02\pm 0.11 R_E$, consistent with the typical dawn-dusk scales of the BBFs in the magnetotail \cite{nakamura_spatial_2004}. This prediction, and the underlying model, are valid under the assumption that the acceleration time must be shorter than the time scale of the magnetic fluctuation or $L_x > L_{x0} = \sqrt{L_y v_{\phi}/\pi f_{cp}}$, where $v_{\phi}$ is the DF velocity assumed to be equal to the bulk velocity and $L_x$ is the spatial scale of the magnetic field fluctuations. It follows that the condition is satisfied $L_x = 0.65\pm 0.08~R_E > L_{x0} = 0.52\pm 0.03~R_E$. Hence, we conclude that the ions energization is due to the resonant interaction of the $\sim 70~\textrm{keV}$ ions with the inductive electric field associated with the DF. 

For the event II (17:17:04-17:22:36 UT), we found that the ion flux changes are well ordered by $K/q$. We note that the charge-state-dependent energization is a necessary condition but it does not necessarily imply adiabaticity of the energization process \cite{catapano_charge_2017,ukhorskiy_ion_2018}. We observed that at energies above the threshold energy of $50~\textrm{keV}$ (i.e., $\rho_p \gtrsim 0.52~R_E$), the energization depends on the initial proton energy. Preferential acceleration of protons with $\rho_p \gtrsim 0.52~R_E$ indicates that the electric field structure (i.e., the acceleration region) is spatially limited. Since the accelerating force is the cross tail motional electric field, the characteristic acceleration scale is thus the dawn-dusk scale $L_y$ of the BBF. The energy dependence of the acceleration above $50~\textrm{keV}$ suggests that the characteristic dawn-dusk scale is comparable to the gyroradius at this threshold energy $\sim 0.52~R_E$.  We conclude that the dawn-dusk scale of the acceleration region is $\sim 0.52~R_E$. This is much smaller than the typical scale of the flow channel $2-3~R_E$ \cite{nakamura_spatial_2004}. 

Using the model of the spatially limited electric field structure proposed by~\citeA{artemyev_acceleration_2015}, we showed that at energies $\sim 100~\textrm{keV}$ the model agrees with our observation while for lower energies, the observed energy gain is lower than the predicted one. This overestimation of the energy gain by the model is likely due to simplified proton orbits in the model proposed by~\citeA{artemyev_acceleration_2015}. Indeed, in the model of spatially limited electric field pulse, the ion is assumed to interact with the dawn-dusk electric field during its complete duskward half gyration. However, this scenario is valid if the ion interacts with the electric field pulse during the duskward half gyration only, and that the dawn-dusk scale of the acceleration region is larger than the ion gyroradius. In particular, this model does not take into account the case of interaction of the ion with the electric field pulse during the dawnward gyration which leads to deceleration of the ion, or the case of an ion which leaves the acceleration before the completing the dawnward half gyration. 

Furthermore, we observed that the energy gained by protons with initial energy $50~\textrm{keV} < K_p < 100~\textrm{keV}$, represents a significant fraction of the initial energy up to $\sim 50~\%$. This result indicates that protons with gyroradius comparable to the dawn-dusk scale $L_y$ of the acceleration region (i.e., dawn-dusk scale of the BBF), gain energy from the motional electric field in a non-adiabatic manner. On the other hand, as mentioned by \cite{artemyev_acceleration_2015}, for protons with gyroradius $\rho_p \gg L_y$ of the electric field structure, the approximation $L_y \sim 2\rho_p$ used in the model is no longer valid, so that the energy increase becomes $\delta K = eE_y L_y$. This means that the energy gain does not depend on the initial energy, and hence $\delta K / K_0 \propto 1 / K_0$ or $\delta \rho_i / \rho_{i0} \propto 1 / \sqrt{K_0}$. For protons with high initial energy, the energy gain during interaction with individual electric pulses, represents only a small fraction of their initial energy, thus their behavior is close to adiabatic. Hence, we conclude that protons with gyroradius comparable with the dawn-dusk scale of the electric field pulse $\rho_p\sim L_y$ are non-adiabatically accelerated, whereas protons with gyroradius much larger $\rho_p >> L_y$ are accelerated in an adiabatic manner.

We showed that the energization occurs for protons with gyroradius comparable with the scale of the acceleration region. So that, $L \sim \rho_i= \sqrt{\alpha} \rho_{H^+} / q$, where $\alpha=m_i / m_p$, and the corresponding ion energy $K_0 = q^2 K_{0, H^+} / \alpha$. In particular, for He$^{2+}$ $\alpha = 4$ and $q=2$, which yields that H$^{+}$ and He$^{2+}$ ions with equal energies $K_{0, He^{2+}} = K_{0, H^{+}}$ have equal gyroradii. However, we showed that, especially for the event II (17:17:04-17:22:36 UT), the flux changes are well ordered by $K/q$ similar to that found in \citeA{catapano_charge_2017} and \citeA{ukhorskiy_ion_2018}. This result suggests that the  acceleration of heavier ions occurs at scales larger than that for protons. On the other hand, preferential acceleration of heavier ions ($m_i/m_p > 1$) may results from interaction with electromagnetic fluctuations with time scales on the heavier ion gyro-periods \cite{keika_energization_2013}. In such a scenario,  the protons with gyroradius smaller than the scale of the structure have a zero net energy gain. At the same time, heavier ions with gyroradius smaller than the scale of the structure but gyroperiod on the time scale of the fluctuations are accelerated through interaction with the inductive electric field.

To summarize, we showed that in both cases the acceleration is due to Earthward moving electric field pulses. The protons with gyroradius smaller than the dawn-dusk scale of the electric field pulse have no net energy gain due to compensation of the energy gain during the duskward half of the gyration by the energy loss during the dawnward half of the gyration. We showed that when the temporal scale of the electromagnetic fluctuations is much larger than the proton gyroperiod (second time interval), protons with gyroradius larger than the scale of the electric field pulse gain energy through crossing of the electric field pulse. On the other hand, when the temporal scale of the electromagnetic fluctuations is of the order of the proton gyroperiod (first time interval) trapped protons with gyroradius of the order of the scale of the electric field pulse gain energy through resonant interaction with the inductive electric field. We note that the non-adiabatic acceleration by spatially limited electric field pulses is much more efficient than the non-adiabatic acceleration by resonant interaction with the inductive electric field.

\section{Conclusion}
\label{sec:conclusion}
We have presented observations of ion acceleration related to plasma jets (bursty bulk flows) in the Earth's magnetotail during a moderate substorm activity. We observed that the $> 100~\textrm{keV}$ ions consist primarily of solar wind Helium He$^{2+}$, while the $< 100~\textrm{keV}$ ion flux is dominated by the protons.

We study in detail two jets fronts with different characteristics. We find that in the event I (16:54:14-16:57:25 UT) with a chain of small-scale magnetic field structures, protons with gyroradius comparable to the scale of the structure gain energy through resonant acceleration \cite{artemyev_ion_2012}. We showed that the interaction of the protons with electromagnetic fluctuations on proton scales leads to a gain of energy in a non-adiabatic manner.

For the event II (17:17:04-17:22:36 UT) with a larger scale magnetic field structure, the protons with a gyroradius larger than scale of the structure, gain energy via interaction with the cross-tail motional electric field during the duskward part of their orbit. We find that the energy gain for protons with gyroradius of the order of the scale of the structure is proportional to their gyroradius and represent a significant fraction of their initial energy. 

In both cases the energization occurs for protons with gyroradii of the order of the scale of the structure or larger. Protons with smaller gyroradii have no net energy gain from the electric field, so that their energy gain in the spacecraft frame is due the earthward convection of the magnetic structure.

\acknowledgments
We thank the entire MMS team and instrument PIs for data access and support. All data used in this paper are publicly available from the MMS Science Data Center \url{https://lasp.colorado.edu/mms/sdc/}. We also wish to thank A. Lalti and the International Space Science Institute (ISSI) working group on "Magnetotail Dipolarizations: Archimedes Force or Ideal Collapse?" for valuable discussions. Data analysis was performed using the pyrfu analysis package available at \url{https://pypi.org/project/pyrfu/}. The codes to reproduce the figures in this paper are available at \url{https://github.com/louis-richard/ionacc} and additional data are available at \url{https://doi.org/10.5281/zenodo.6320624}. This work is supported by the Swedish National Space Agency grant 139/18.

\bibliography{main.bib}

\begin{thebibliography}{}

\bibitem [\protect \citeauthoryear {%
Angelopoulos%
\ \protect \BOthers {.}}{%
Angelopoulos%
\ \protect \BOthers {.}}{%
{\protect \APACyear {1992}}%
}]{%
angelopoulos_bursty_1992}
\APACinsertmetastar {%
angelopoulos_bursty_1992}%
\begin{APACrefauthors}%
Angelopoulos, V.%
, Baumjohann, W.%
, Kennel, C\BPBI F.%
, Coroniti, F\BPBI V.%
, Kivelson, M\BPBI G.%
, Pellat, R.%
\BDBL {}Paschmann, G.%
\end{APACrefauthors}%
\unskip\
\newblock
\APACrefYearMonthDay{1992}{}{}.
\newblock
{\BBOQ}\APACrefatitle {Bursty bulk flows in the inner central plasma sheet}
  {Bursty bulk flows in the inner central plasma sheet}.{\BBCQ}
\newblock
\APACjournalVolNumPages{Journal of Geophysical Research}{97}{A4}{4027--4039}.
\newblock
\begin{APACrefDOI} \doi{10.1029/91JA02701} \end{APACrefDOI}
\PrintBackRefs{\CurrentBib}

\bibitem [\protect \citeauthoryear {%
Angelopoulos%
\ \protect \BOthers {.}}{%
Angelopoulos%
\ \protect \BOthers {.}}{%
{\protect \APACyear {1994}}%
}]{%
angelopoulos_statistical_1994}
\APACinsertmetastar {%
angelopoulos_statistical_1994}%
\begin{APACrefauthors}%
Angelopoulos, V.%
, Kennel, C\BPBI F.%
, Coroniti, F\BPBI V.%
, Pellat, R.%
, Kivelson, M\BPBI G.%
, Walker, R\BPBI J.%
\BDBL {}Gosling, J\BPBI T.%
\end{APACrefauthors}%
\unskip\
\newblock
\APACrefYearMonthDay{1994}{}{}.
\newblock
{\BBOQ}\APACrefatitle {Statistical characteristics of bursty bulk flow events}
  {Statistical characteristics of bursty bulk flow events}.{\BBCQ}
\newblock
\APACjournalVolNumPages{Journal of Geophysical
  Research}{99}{A11}{21257--21280}.
\newblock
\begin{APACrefDOI} \doi{10.1029/94JA01263} \end{APACrefDOI}
\PrintBackRefs{\CurrentBib}

\bibitem [\protect \citeauthoryear {%
Apatenkov%
\ \protect \BOthers {.}}{%
Apatenkov%
\ \protect \BOthers {.}}{%
{\protect \APACyear {2007}}%
}]{%
apatenkov_multi-spacecraft_2007}
\APACinsertmetastar {%
apatenkov_multi-spacecraft_2007}%
\begin{APACrefauthors}%
Apatenkov, S\BPBI V.%
, Sergeev, V\BPBI A.%
, Kubyshkina, M\BPBI V.%
, Nakamura, R.%
, Baumjohann, W.%
, Runov, A.%
\BDBL {}Khotyaintsev, Y.%
\end{APACrefauthors}%
\unskip\
\newblock
\APACrefYearMonthDay{2007}{}{}.
\newblock
{\BBOQ}\APACrefatitle {Multi-spacecraft observation of plasma
  dipolarization/injection in the inner magnetosphere} {Multi-spacecraft
  observation of plasma dipolarization/injection in the inner
  magnetosphere}.{\BBCQ}
\newblock
\APACjournalVolNumPages{Annales Geophysicae}{25}{3}{801--814}.
\newblock
\begin{APACrefDOI} \doi{10.5194/angeo-25-801-2007} \end{APACrefDOI}
\PrintBackRefs{\CurrentBib}

\bibitem [\protect \citeauthoryear {%
Artemyev%
, Liu%
, Angelopoulos%
\BCBL {}\ \BBA {} Runov%
}{%
Artemyev%
\ \protect \BOthers {.}}{%
{\protect \APACyear {2015}}%
}]{%
artemyev_acceleration_2015}
\APACinsertmetastar {%
artemyev_acceleration_2015}%
\begin{APACrefauthors}%
Artemyev, A\BPBI V.%
, Liu, J.%
, Angelopoulos, V.%
\BCBL {}\ \BBA {} Runov, A.%
\end{APACrefauthors}%
\unskip\
\newblock
\APACrefYearMonthDay{2015}{}{}.
\newblock
{\BBOQ}\APACrefatitle {Acceleration of ions by electric field pulses in the
  inner magnetosphere} {Acceleration of ions by electric field pulses in the
  inner magnetosphere}.{\BBCQ}
\newblock
\APACjournalVolNumPages{Journal of Geophysical Research: Space
  Physics}{120}{6}{4628--4640}.
\newblock
\begin{APACrefDOI} \doi{10.1002/2015JA021160} \end{APACrefDOI}
\PrintBackRefs{\CurrentBib}

\bibitem [\protect \citeauthoryear {%
Artemyev%
, Lutsenko%
\BCBL {}\ \BBA {} Petrukovich%
}{%
Artemyev%
\ \protect \BOthers {.}}{%
{\protect \APACyear {2012}}%
}]{%
artemyev_ion_2012}
\APACinsertmetastar {%
artemyev_ion_2012}%
\begin{APACrefauthors}%
Artemyev, A\BPBI V.%
, Lutsenko, V\BPBI N.%
\BCBL {}\ \BBA {} Petrukovich, A\BPBI A.%
\end{APACrefauthors}%
\unskip\
\newblock
\APACrefYearMonthDay{2012}{}{}.
\newblock
{\BBOQ}\APACrefatitle {Ion resonance acceleration by dipolarization fronts:
  analytic theory and spacecraft observation} {Ion resonance acceleration by
  dipolarization fronts: analytic theory and spacecraft observation}.{\BBCQ}
\newblock
\APACjournalVolNumPages{Annales Geophysicae}{30}{2}{317--324}.
\newblock
\begin{APACrefDOI} \doi{10.5194/angeo-30-317-2012} \end{APACrefDOI}
\PrintBackRefs{\CurrentBib}

\bibitem [\protect \citeauthoryear {%
Baumjohann%
, Paschmann%
\BCBL {}\ \BBA {} Lühr%
}{%
Baumjohann%
\ \protect \BOthers {.}}{%
{\protect \APACyear {1990}}%
}]{%
baumjohann_characteristics_1990}
\APACinsertmetastar {%
baumjohann_characteristics_1990}%
\begin{APACrefauthors}%
Baumjohann, W.%
, Paschmann, G.%
\BCBL {}\ \BBA {} Lühr, H.%
\end{APACrefauthors}%
\unskip\
\newblock
\APACrefYearMonthDay{1990}{}{}.
\newblock
{\BBOQ}\APACrefatitle {Characteristics of high-speed ion flows in the plasma
  sheet} {Characteristics of high-speed ion flows in the plasma sheet}.{\BBCQ}
\newblock
\APACjournalVolNumPages{Journal of Geophysical Research}{95}{A4}{3801--3809}.
\newblock
\begin{APACrefDOI} \doi{10.1029/JA095iA04p03801} \end{APACrefDOI}
\PrintBackRefs{\CurrentBib}

\bibitem [\protect \citeauthoryear {%
Bingham%
\ \protect \BOthers {.}}{%
Bingham%
\ \protect \BOthers {.}}{%
{\protect \APACyear {2020}}%
}]{%
bingham_chargestatedependent_2020}
\APACinsertmetastar {%
bingham_chargestatedependent_2020}%
\begin{APACrefauthors}%
Bingham, S\BPBI T.%
, Cohen, I\BPBI J.%
, Mauk, B\BPBI H.%
, Turner, D\BPBI L.%
, Mitchell, D\BPBI G.%
, Vines, S\BPBI K.%
\BDBL {}Burch, J\BPBI L.%
\end{APACrefauthors}%
\unskip\
\newblock
\APACrefYearMonthDay{2020}{}{}.
\newblock
{\BBOQ}\APACrefatitle {Charge‐state‐dependent energization of suprathermal
  ions during substorm injections observed by {MMS} in the magnetotail}
  {Charge‐state‐dependent energization of suprathermal ions during substorm
  injections observed by {MMS} in the magnetotail}.{\BBCQ}
\newblock
\APACjournalVolNumPages{Journal of Geophysical Research: Space
  Physics}{125}{}{}.
\newblock
\begin{APACrefDOI} \doi{10.1029/2020JA028144} \end{APACrefDOI}
\PrintBackRefs{\CurrentBib}

\bibitem [\protect \citeauthoryear {%
Bingham%
\ \protect \BOthers {.}}{%
Bingham%
\ \protect \BOthers {.}}{%
{\protect \APACyear {2021}}%
}]{%
bingham_evidence_2021}
\APACinsertmetastar {%
bingham_evidence_2021}%
\begin{APACrefauthors}%
Bingham, S\BPBI T.%
, Nikoukar, R.%
, Cohen, I\BPBI J.%
, Mauk, B\BPBI H.%
, Turner, D\BPBI L.%
, Mitchell, D\BPBI G.%
\BDBL {}Torbert, R\BPBI B.%
\end{APACrefauthors}%
\unskip\
\newblock
\APACrefYearMonthDay{2021}{}{}.
\newblock
{\BBOQ}\APACrefatitle {Evidence for nonadiabatic oxygen energization in the
  near‐{Earth} magnetotail from {MMS}} {Evidence for nonadiabatic oxygen
  energization in the near‐{Earth} magnetotail from {MMS}}.{\BBCQ}
\newblock
\APACjournalVolNumPages{Geophysical Research Letters}{48}{}{}.
\newblock
\begin{APACrefDOI} \doi{10.1029/2020GL091697} \end{APACrefDOI}
\PrintBackRefs{\CurrentBib}

\bibitem [\protect \citeauthoryear {%
Birn%
\ \BBA {} Hesse%
}{%
Birn%
\ \BBA {} Hesse%
}{%
{\protect \APACyear {1994}}%
}]{%
birn_particle_1994}
\APACinsertmetastar {%
birn_particle_1994}%
\begin{APACrefauthors}%
Birn, J.%
\BCBT {}\ \BBA {} Hesse, M.%
\end{APACrefauthors}%
\unskip\
\newblock
\APACrefYearMonthDay{1994}{}{}.
\newblock
{\BBOQ}\APACrefatitle {Particle acceleration in the dynamic magnetotail:
  {Orbits} in self-consistent three-dimensional {MHD} fields} {Particle
  acceleration in the dynamic magnetotail: {Orbits} in self-consistent
  three-dimensional {MHD} fields}.{\BBCQ}
\newblock
\APACjournalVolNumPages{Journal of Geophysical Research}{99}{A1}{109}.
\newblock
\begin{APACrefDOI} \doi{10.1029/93JA02284} \end{APACrefDOI}
\PrintBackRefs{\CurrentBib}

\bibitem [\protect \citeauthoryear {%
Birn%
, Hesse%
, Nakamura%
\BCBL {}\ \BBA {} Zaharia%
}{%
Birn%
\ \protect \BOthers {.}}{%
{\protect \APACyear {2013}}%
}]{%
birn_particle_2013}
\APACinsertmetastar {%
birn_particle_2013}%
\begin{APACrefauthors}%
Birn, J.%
, Hesse, M.%
, Nakamura, R.%
\BCBL {}\ \BBA {} Zaharia, S.%
\end{APACrefauthors}%
\unskip\
\newblock
\APACrefYearMonthDay{2013}{}{}.
\newblock
{\BBOQ}\APACrefatitle {Particle acceleration in dipolarization events}
  {Particle acceleration in dipolarization events}.{\BBCQ}
\newblock
\APACjournalVolNumPages{Journal of Geophysical Research: Space
  Physics}{118}{5}{1960--1971}.
\newblock
\begin{APACrefDOI} \doi{10.1002/jgra.50132} \end{APACrefDOI}
\PrintBackRefs{\CurrentBib}

\bibitem [\protect \citeauthoryear {%
Birn%
, Runov%
\BCBL {}\ \BBA {} Hesse%
}{%
Birn%
\ \protect \BOthers {.}}{%
{\protect \APACyear {2015}}%
}]{%
birn_energetic_2015}
\APACinsertmetastar {%
birn_energetic_2015}%
\begin{APACrefauthors}%
Birn, J.%
, Runov, A.%
\BCBL {}\ \BBA {} Hesse, M.%
\end{APACrefauthors}%
\unskip\
\newblock
\APACrefYearMonthDay{2015}{}{}.
\newblock
{\BBOQ}\APACrefatitle {Energetic ions in dipolarization events} {Energetic ions
  in dipolarization events}.{\BBCQ}
\newblock
\APACjournalVolNumPages{Journal of Geophysical Research: Space
  Physics}{120}{9}{7698--7717}.
\newblock
\begin{APACrefDOI} \doi{10.1002/2015JA021372} \end{APACrefDOI}
\PrintBackRefs{\CurrentBib}

\bibitem [\protect \citeauthoryear {%
Birn%
, Runov%
\BCBL {}\ \BBA {} Zhou%
}{%
Birn%
\ \protect \BOthers {.}}{%
{\protect \APACyear {2017}}%
}]{%
birn_ion_2017}
\APACinsertmetastar {%
birn_ion_2017}%
\begin{APACrefauthors}%
Birn, J.%
, Runov, A.%
\BCBL {}\ \BBA {} Zhou, X.%
\end{APACrefauthors}%
\unskip\
\newblock
\APACrefYearMonthDay{2017}{}{}.
\newblock
{\BBOQ}\APACrefatitle {Ion velocity distributions in dipolarization events:
  {Distributions} in the central plasma sheet} {Ion velocity distributions in
  dipolarization events: {Distributions} in the central plasma sheet}.{\BBCQ}
\newblock
\APACjournalVolNumPages{Journal of Geophysical Research: Space
  Physics}{122}{8}{8014--8025}.
\newblock
\begin{APACrefDOI} \doi{10.1002/2017JA024230} \end{APACrefDOI}
\PrintBackRefs{\CurrentBib}

\bibitem [\protect \citeauthoryear {%
Birn%
, Thomsen%
, Borovsky%
, Reeves%
\BCBL {}\ \BBA {} Hesse%
}{%
Birn%
\ \protect \BOthers {.}}{%
{\protect \APACyear {2000}}%
}]{%
birn_particle_2000}
\APACinsertmetastar {%
birn_particle_2000}%
\begin{APACrefauthors}%
Birn, J.%
, Thomsen, M\BPBI F.%
, Borovsky, J\BPBI E.%
, Reeves, G\BPBI D.%
\BCBL {}\ \BBA {} Hesse, M.%
\end{APACrefauthors}%
\unskip\
\newblock
\APACrefYearMonthDay{2000}{}{}.
\newblock
{\BBOQ}\APACrefatitle {Particle acceleration in the dynamic magnetotail}
  {Particle acceleration in the dynamic magnetotail}.{\BBCQ}
\newblock
\APACjournalVolNumPages{Physics of Plasmas}{7}{5}{2149--2156}.
\newblock
\begin{APACrefDOI} \doi{10.1063/1.874035} \end{APACrefDOI}
\PrintBackRefs{\CurrentBib}

\bibitem [\protect \citeauthoryear {%
Birn%
\ \protect \BOthers {.}}{%
Birn%
\ \protect \BOthers {.}}{%
{\protect \APACyear {1997}}%
}]{%
birn_substorm_1997}
\APACinsertmetastar {%
birn_substorm_1997}%
\begin{APACrefauthors}%
Birn, J.%
, Thomsen, M\BPBI F.%
, Borovsky, J\BPBI E.%
, Reeves, G\BPBI D.%
, McComas, D\BPBI J.%
, Belian, R\BPBI D.%
\BCBL {}\ \BBA {} Hesse, M.%
\end{APACrefauthors}%
\unskip\
\newblock
\APACrefYearMonthDay{1997}{}{}.
\newblock
{\BBOQ}\APACrefatitle {Substorm ion injections: {Geosynchronous} observations
  and test particle orbits in three-dimensional dynamic {MHD} fields} {Substorm
  ion injections: {Geosynchronous} observations and test particle orbits in
  three-dimensional dynamic {MHD} fields}.{\BBCQ}
\newblock
\APACjournalVolNumPages{Journal of Geophysical Research: Space
  Physics}{102}{A2}{2325--2341}.
\newblock
\begin{APACrefDOI} \doi{10.1029/96JA03032} \end{APACrefDOI}
\PrintBackRefs{\CurrentBib}

\bibitem [\protect \citeauthoryear {%
Blake%
\ \protect \BOthers {.}}{%
Blake%
\ \protect \BOthers {.}}{%
{\protect \APACyear {2016}}%
}]{%
blake_flys_2016}
\APACinsertmetastar {%
blake_flys_2016}%
\begin{APACrefauthors}%
Blake, J\BPBI B.%
, Mauk, B\BPBI H.%
, Baker, D\BPBI N.%
, Carranza, P.%
, Clemmons, J\BPBI H.%
, Craft, J.%
\BDBL {}Westlake, J.%
\end{APACrefauthors}%
\unskip\
\newblock
\APACrefYearMonthDay{2016}{}{}.
\newblock
{\BBOQ}\APACrefatitle {The {Fly}’s {Eye} {Energetic} {Particle}
  {Spectrometer} ({FEEPS}) {Sensors} for the {Magnetospheric} {Multiscale}
  ({MMS}) {Mission}} {The {Fly}’s {Eye} {Energetic} {Particle} {Spectrometer}
  ({FEEPS}) {Sensors} for the {Magnetospheric} {Multiscale} ({MMS})
  {Mission}}.{\BBCQ}
\newblock
\APACjournalVolNumPages{Space Science Reviews}{199}{1-4}{309--329}.
\newblock
\begin{APACrefDOI} \doi{10.1007/s11214-015-0163-x} \end{APACrefDOI}
\PrintBackRefs{\CurrentBib}

\bibitem [\protect \citeauthoryear {%
Catapano%
\ \protect \BOthers {.}}{%
Catapano%
\ \protect \BOthers {.}}{%
{\protect \APACyear {2017}}%
}]{%
catapano_charge_2017}
\APACinsertmetastar {%
catapano_charge_2017}%
\begin{APACrefauthors}%
Catapano, F.%
, Zimbardo, G.%
, Perri, S.%
, Greco, A.%
, Delcourt, D.%
, Retinò, A.%
\BCBL {}\ \BBA {} Cohen, I\BPBI J.%
\end{APACrefauthors}%
\unskip\
\newblock
\APACrefYearMonthDay{2017}{}{}.
\newblock
{\BBOQ}\APACrefatitle {Charge {Proportional} and {Weakly} {Mass}-{Dependent}
  {Acceleration} of {Different} {Ion} {Species} in the {Earth}'s {Magnetotail}}
  {Charge {Proportional} and {Weakly} {Mass}-{Dependent} {Acceleration} of
  {Different} {Ion} {Species} in the {Earth}'s {Magnetotail}}.{\BBCQ}
\newblock
\APACjournalVolNumPages{Geophysical Research Letters}{44}{}{10108--10115}.
\newblock
\begin{APACrefDOI} \doi{10.1002/2017GL075092} \end{APACrefDOI}
\PrintBackRefs{\CurrentBib}

\bibitem [\protect \citeauthoryear {%
Chen%
\ \BBA {} Wolf%
}{%
Chen%
\ \BBA {} Wolf%
}{%
{\protect \APACyear {1999}}%
}]{%
chen_theory_1999}
\APACinsertmetastar {%
chen_theory_1999}%
\begin{APACrefauthors}%
Chen, C\BPBI X.%
\BCBT {}\ \BBA {} Wolf, R\BPBI A.%
\end{APACrefauthors}%
\unskip\
\newblock
\APACrefYearMonthDay{1999}{}{}.
\newblock
{\BBOQ}\APACrefatitle {Theory of thin-filament motion in {Earth}'s magnetotail
  and its application to bursty bulk flows} {Theory of thin-filament motion in
  {Earth}'s magnetotail and its application to bursty bulk flows}.{\BBCQ}
\newblock
\APACjournalVolNumPages{Journal of Geophysical Research: Space
  Physics}{104}{A7}{14613--14626}.
\newblock
\begin{APACrefDOI} \doi{10.1029/1999JA900005} \end{APACrefDOI}
\PrintBackRefs{\CurrentBib}

\bibitem [\protect \citeauthoryear {%
Cohen%
\ \protect \BOthers {.}}{%
Cohen%
\ \protect \BOthers {.}}{%
{\protect \APACyear {2017}}%
}]{%
cohen_dominance_2017}
\APACinsertmetastar {%
cohen_dominance_2017}%
\begin{APACrefauthors}%
Cohen, I\BPBI J.%
, Mitchell, D\BPBI G.%
, Kistler, L\BPBI M.%
, Mauk, B\BPBI H.%
, Anderson, B\BPBI J.%
, Westlake, J\BPBI H.%
\BDBL {}Burch, J\BPBI L.%
\end{APACrefauthors}%
\unskip\
\newblock
\APACrefYearMonthDay{2017}{}{}.
\newblock
{\BBOQ}\APACrefatitle {Dominance of high-energy ({\textgreater}150 {keV}) heavy
  ion intensities in {Earth}'s middle to outer magnetosphere} {Dominance of
  high-energy ({\textgreater}150 {keV}) heavy ion intensities in {Earth}'s
  middle to outer magnetosphere}.{\BBCQ}
\newblock
\APACjournalVolNumPages{Journal of Geophysical Research: Space
  Physics}{122}{9}{9282--9293}.
\newblock
\begin{APACrefDOI} \doi{10.1002/2017JA024351} \end{APACrefDOI}
\PrintBackRefs{\CurrentBib}

\bibitem [\protect \citeauthoryear {%
Ergun%
\ \protect \BOthers {.}}{%
Ergun%
\ \protect \BOthers {.}}{%
{\protect \APACyear {2016}}%
}]{%
ergun_axial_2016}
\APACinsertmetastar {%
ergun_axial_2016}%
\begin{APACrefauthors}%
Ergun, R\BPBI E.%
, Tucker, S.%
, Westfall, J.%
, Goodrich, K\BPBI A.%
, Malaspina, D\BPBI M.%
, Summers, D.%
\BDBL {}Cully, C\BPBI M.%
\end{APACrefauthors}%
\unskip\
\newblock
\APACrefYearMonthDay{2016}{}{}.
\newblock
{\BBOQ}\APACrefatitle {The {Axial} {Double} {Probe} and {Fields} {Signal}
  {Processing} for the {MMS} {Mission}} {The {Axial} {Double} {Probe} and
  {Fields} {Signal} {Processing} for the {MMS} {Mission}}.{\BBCQ}
\newblock
\APACjournalVolNumPages{Space Science Reviews}{199}{1-4}{167--188}.
\newblock
\begin{APACrefDOI} \doi{10.1007/s11214-014-0115-x} \end{APACrefDOI}
\PrintBackRefs{\CurrentBib}

\bibitem [\protect \citeauthoryear {%
Fu%
\ \protect \BOthers {.}}{%
Fu%
\ \protect \BOthers {.}}{%
{\protect \APACyear {2020}}%
}]{%
fu_magnetotail_2020}
\APACinsertmetastar {%
fu_magnetotail_2020}%
\begin{APACrefauthors}%
Fu, H\BPBI S.%
, Grigorenko, E\BPBI E.%
, Gabrielse, C.%
, Liu, C.%
, Lu, S.%
, Hwang, K\BPBI J.%
\BDBL {}Chen, F.%
\end{APACrefauthors}%
\unskip\
\newblock
\APACrefYearMonthDay{2020}{}{}.
\newblock
{\BBOQ}\APACrefatitle {Magnetotail dipolarization fronts and particle
  acceleration: {A} review} {Magnetotail dipolarization fronts and particle
  acceleration: {A} review}.{\BBCQ}
\newblock
\APACjournalVolNumPages{Science China Earth Sciences}{63}{2}{235--256}.
\newblock
\begin{APACrefDOI} \doi{10.1007/s11430-019-9551-y} \end{APACrefDOI}
\PrintBackRefs{\CurrentBib}

\bibitem [\protect \citeauthoryear {%
Fu%
, Khotyaintsev%
, André%
\BCBL {}\ \BBA {} Vaivads%
}{%
Fu%
\ \protect \BOthers {.}}{%
{\protect \APACyear {2011}}%
}]{%
fu_fermi_2011}
\APACinsertmetastar {%
fu_fermi_2011}%
\begin{APACrefauthors}%
Fu, H\BPBI S.%
, Khotyaintsev, Y\BPBI V.%
, André, M.%
\BCBL {}\ \BBA {} Vaivads, A.%
\end{APACrefauthors}%
\unskip\
\newblock
\APACrefYearMonthDay{2011}{}{}.
\newblock
{\BBOQ}\APACrefatitle {Fermi and betatron acceleration of suprathermal
  electrons behind dipolarization fronts} {Fermi and betatron acceleration of
  suprathermal electrons behind dipolarization fronts}.{\BBCQ}
\newblock
\APACjournalVolNumPages{Geophysical Research Letters}{38}{}{L16104}.
\newblock
\begin{APACrefDOI} \doi{10.1029/2011GL048528} \end{APACrefDOI}
\PrintBackRefs{\CurrentBib}

\bibitem [\protect \citeauthoryear {%
Fu%
, Khotyaintsev%
, Vaivads%
, Retinò%
\BCBL {}\ \BBA {} André%
}{%
Fu%
\ \protect \BOthers {.}}{%
{\protect \APACyear {2013}}%
}]{%
fu_energetic_2013}
\APACinsertmetastar {%
fu_energetic_2013}%
\begin{APACrefauthors}%
Fu, H\BPBI S.%
, Khotyaintsev, Y\BPBI V.%
, Vaivads, A.%
, Retinò, A.%
\BCBL {}\ \BBA {} André, M.%
\end{APACrefauthors}%
\unskip\
\newblock
\APACrefYearMonthDay{2013}{}{}.
\newblock
{\BBOQ}\APACrefatitle {Energetic electron acceleration by unsteady magnetic
  reconnection} {Energetic electron acceleration by unsteady magnetic
  reconnection}.{\BBCQ}
\newblock
\APACjournalVolNumPages{Nature Physics}{9}{7}{426--430}.
\newblock
\begin{APACrefDOI} \doi{10.1038/nphys2664} \end{APACrefDOI}
\PrintBackRefs{\CurrentBib}

\bibitem [\protect \citeauthoryear {%
Gershman%
\ \protect \BOthers {.}}{%
Gershman%
\ \protect \BOthers {.}}{%
{\protect \APACyear {2019}}%
}]{%
gershman_systematic_2019}
\APACinsertmetastar {%
gershman_systematic_2019}%
\begin{APACrefauthors}%
Gershman, D\BPBI J.%
, Dorelli, J\BPBI C.%
, Avanov, L\BPBI A.%
, Gliese, U.%
, Barrie, A.%
, Schiff, C.%
\BDBL {}Pollock, C\BPBI J.%
\end{APACrefauthors}%
\unskip\
\newblock
\APACrefYearMonthDay{2019}{}{}.
\newblock
{\BBOQ}\APACrefatitle {Systematic uncertainties in plasma parameters reported
  by the {Fast} {Plasma} {Investigation} on {NASA}'s {Magnetospheric}
  {Multiscale} mission} {Systematic uncertainties in plasma parameters reported
  by the {Fast} {Plasma} {Investigation} on {NASA}'s {Magnetospheric}
  {Multiscale} mission}.{\BBCQ}
\newblock
\APACjournalVolNumPages{Journal of Geophysical Research: Space
  Physics}{124}{12}{10345--10359}.
\newblock
\begin{APACrefDOI} \doi{10.1029/2019JA026980} \end{APACrefDOI}
\PrintBackRefs{\CurrentBib}

\bibitem [\protect \citeauthoryear {%
Gosling%
\ \protect \BOthers {.}}{%
Gosling%
\ \protect \BOthers {.}}{%
{\protect \APACyear {1982}}%
}]{%
gosling_evidence_1982}
\APACinsertmetastar {%
gosling_evidence_1982}%
\begin{APACrefauthors}%
Gosling, J\BPBI T.%
, Thomsen, M\BPBI F.%
, Bame, S\BPBI J.%
, Feldman, W\BPBI C.%
, Paschmann, G.%
\BCBL {}\ \BBA {} Sckopke, N.%
\end{APACrefauthors}%
\unskip\
\newblock
\APACrefYearMonthDay{1982}{}{}.
\newblock
{\BBOQ}\APACrefatitle {Evidence for specularly reflected ions upstream from the
  quasi-parallel bow shock} {Evidence for specularly reflected ions upstream
  from the quasi-parallel bow shock}.{\BBCQ}
\newblock
\APACjournalVolNumPages{Geophysical Research Letters}{9}{12}{1333--1336}.
\newblock
\begin{APACrefDOI} \doi{10.1029/GL009i012p01333} \end{APACrefDOI}
\PrintBackRefs{\CurrentBib}

\bibitem [\protect \citeauthoryear {%
Haaland%
\ \protect \BOthers {.}}{%
Haaland%
\ \protect \BOthers {.}}{%
{\protect \APACyear {2010}}%
}]{%
haaland_spectral_2010}
\APACinsertmetastar {%
haaland_spectral_2010}%
\begin{APACrefauthors}%
Haaland, S.%
, Kronberg, E\BPBI A.%
, Daly, P\BPBI W.%
, Fränz, M.%
, Degener, L.%
, Georgescu, E.%
\BCBL {}\ \BBA {} Dandouras, I.%
\end{APACrefauthors}%
\unskip\
\newblock
\APACrefYearMonthDay{2010}{}{}.
\newblock
{\BBOQ}\APACrefatitle {Spectral characteristics of protons in the {Earth}'s
  plasmasheet: statistical results from {Cluster} {CIS} and {RAPID}} {Spectral
  characteristics of protons in the {Earth}'s plasmasheet: statistical results
  from {Cluster} {CIS} and {RAPID}}.{\BBCQ}
\newblock
\APACjournalVolNumPages{Annales Geophysicae}{28}{8}{1483--1498}.
\newblock
\begin{APACrefDOI} \doi{10.5194/angeo-28-1483-2010} \end{APACrefDOI}
\PrintBackRefs{\CurrentBib}

\bibitem [\protect \citeauthoryear {%
Keika%
, Kistler%
\BCBL {}\ \BBA {} Brandt%
}{%
Keika%
\ \protect \BOthers {.}}{%
{\protect \APACyear {2013}}%
}]{%
keika_energization_2013}
\APACinsertmetastar {%
keika_energization_2013}%
\begin{APACrefauthors}%
Keika, K.%
, Kistler, L\BPBI M.%
\BCBL {}\ \BBA {} Brandt, P.%
\end{APACrefauthors}%
\unskip\
\newblock
\APACrefYearMonthDay{2013}{}{}.
\newblock
{\BBOQ}\APACrefatitle {Energization of {O}+ ions in the {Earth}'s inner
  magnetosphere and the effects on ring current buildup: {A} review of previous
  observations and possible mechanisms} {Energization of {O}+ ions in the
  {Earth}'s inner magnetosphere and the effects on ring current buildup: {A}
  review of previous observations and possible mechanisms}.{\BBCQ}
\newblock
\APACjournalVolNumPages{Journal of Geophysical Research: Space
  Physics}{118}{7}{4441--4464}.
\newblock
\begin{APACrefDOI} \doi{10.1002/jgra.50371} \end{APACrefDOI}
\PrintBackRefs{\CurrentBib}

\bibitem [\protect \citeauthoryear {%
Khotyaintsev%
, Cully%
, Vaivads%
, André%
\BCBL {}\ \BBA {} Owen%
}{%
Khotyaintsev%
\ \protect \BOthers {.}}{%
{\protect \APACyear {2011}}%
}]{%
khotyaintsev_plasma_2011}
\APACinsertmetastar {%
khotyaintsev_plasma_2011}%
\begin{APACrefauthors}%
Khotyaintsev, Y\BPBI V.%
, Cully, C\BPBI M.%
, Vaivads, A.%
, André, M.%
\BCBL {}\ \BBA {} Owen, C\BPBI J.%
\end{APACrefauthors}%
\unskip\
\newblock
\APACrefYearMonthDay{2011}{}{}.
\newblock
{\BBOQ}\APACrefatitle {Plasma jet braking: {Energy} dissipation and
  nonadiabatic electrons} {Plasma jet braking: {Energy} dissipation and
  nonadiabatic electrons}.{\BBCQ}
\newblock
\APACjournalVolNumPages{Physical Review Letters}{106}{16}{165001}.
\newblock
\begin{APACrefDOI} \doi{10.1103/PhysRevLett.106.165001} \end{APACrefDOI}
\PrintBackRefs{\CurrentBib}

\bibitem [\protect \citeauthoryear {%
Kistler%
\ \protect \BOthers {.}}{%
Kistler%
\ \protect \BOthers {.}}{%
{\protect \APACyear {1989}}%
}]{%
kistler_energy_1989}
\APACinsertmetastar {%
kistler_energy_1989}%
\begin{APACrefauthors}%
Kistler, L\BPBI M.%
, Ipavich, F\BPBI M.%
, Hamilton, D\BPBI C.%
, Gloeckler, G.%
, Wilken, B.%
, Kremser, G.%
\BCBL {}\ \BBA {} Stüdemann, W.%
\end{APACrefauthors}%
\unskip\
\newblock
\APACrefYearMonthDay{1989}{}{}.
\newblock
{\BBOQ}\APACrefatitle {Energy spectra of the major ion species in the ring
  current during geomagnetic storms} {Energy spectra of the major ion species
  in the ring current during geomagnetic storms}.{\BBCQ}
\newblock
\APACjournalVolNumPages{Journal of Geophysical Research}{94}{A4}{3579--3599}.
\newblock
\begin{APACrefDOI} \doi{10.1029/JA094iA04p03579} \end{APACrefDOI}
\PrintBackRefs{\CurrentBib}

\bibitem [\protect \citeauthoryear {%
Lindqvist%
\ \protect \BOthers {.}}{%
Lindqvist%
\ \protect \BOthers {.}}{%
{\protect \APACyear {2016}}%
}]{%
lindqvist_spin-plane_2016}
\APACinsertmetastar {%
lindqvist_spin-plane_2016}%
\begin{APACrefauthors}%
Lindqvist, P\BHBI A.%
, Olsson, G.%
, Torbert, R\BPBI B.%
, King, B.%
, Granoff, M.%
, Rau, D.%
\BDBL {}Tucker, S.%
\end{APACrefauthors}%
\unskip\
\newblock
\APACrefYearMonthDay{2016}{}{}.
\newblock
{\BBOQ}\APACrefatitle {The {Spin}-{Plane} {Double} {Probe} {Electric} {Field}
  {Instrument} for {MMS}} {The {Spin}-{Plane} {Double} {Probe} {Electric}
  {Field} {Instrument} for {MMS}}.{\BBCQ}
\newblock
\APACjournalVolNumPages{Space Science Reviews}{199}{1-4}{137--165}.
\newblock
\begin{APACrefDOI} \doi{10.1007/s11214-014-0116-9} \end{APACrefDOI}
\PrintBackRefs{\CurrentBib}

\bibitem [\protect \citeauthoryear {%
Liu%
, Angelopoulos%
, Runov%
\BCBL {}\ \BBA {} Zhou%
}{%
Liu%
\ \protect \BOthers {.}}{%
{\protect \APACyear {2013}}%
}]{%
liu_current_2013}
\APACinsertmetastar {%
liu_current_2013}%
\begin{APACrefauthors}%
Liu, J.%
, Angelopoulos, V.%
, Runov, A.%
\BCBL {}\ \BBA {} Zhou, X\BHBI Z.%
\end{APACrefauthors}%
\unskip\
\newblock
\APACrefYearMonthDay{2013}{}{}.
\newblock
{\BBOQ}\APACrefatitle {On the current sheets surrounding dipolarizing flux
  bundles in the magnetotail: {The} case for wedgelets} {On the current sheets
  surrounding dipolarizing flux bundles in the magnetotail: {The} case for
  wedgelets}.{\BBCQ}
\newblock
\APACjournalVolNumPages{Journal of Geophysical Research: Space
  Physics}{118}{5}{2000--2020}.
\newblock
\begin{APACrefDOI} \doi{10.1002/jgra.50092} \end{APACrefDOI}
\PrintBackRefs{\CurrentBib}

\bibitem [\protect \citeauthoryear {%
Malykhin%
\ \protect \BOthers {.}}{%
Malykhin%
\ \protect \BOthers {.}}{%
{\protect \APACyear {2018}}%
}]{%
malykhin_contrasting_2018}
\APACinsertmetastar {%
malykhin_contrasting_2018}%
\begin{APACrefauthors}%
Malykhin, A\BPBI Y.%
, Grigorenko, E\BPBI E.%
, Kronberg, E\BPBI A.%
, Koleva, R.%
, Ganushkina, N\BPBI Y.%
, Kozak, L.%
\BCBL {}\ \BBA {} Daly, P\BPBI W.%
\end{APACrefauthors}%
\unskip\
\newblock
\APACrefYearMonthDay{2018}{}{}.
\newblock
{\BBOQ}\APACrefatitle {Contrasting dynamics of electrons and protons in the
  near-{Earth} plasma sheet during dipolarization} {Contrasting dynamics of
  electrons and protons in the near-{Earth} plasma sheet during
  dipolarization}.{\BBCQ}
\newblock
\APACjournalVolNumPages{Annales Geophysicae}{36}{3}{741--760}.
\newblock
\begin{APACrefDOI} \doi{10.5194/angeo-36-741-2018} \end{APACrefDOI}
\PrintBackRefs{\CurrentBib}

\bibitem [\protect \citeauthoryear {%
Masuda%
, Kosugi%
, Hara%
, Tsuneta%
\BCBL {}\ \BBA {} Ogawara%
}{%
Masuda%
\ \protect \BOthers {.}}{%
{\protect \APACyear {1994}}%
}]{%
masuda_loop-top_1994}
\APACinsertmetastar {%
masuda_loop-top_1994}%
\begin{APACrefauthors}%
Masuda, S.%
, Kosugi, T.%
, Hara, H.%
, Tsuneta, S.%
\BCBL {}\ \BBA {} Ogawara, Y.%
\end{APACrefauthors}%
\unskip\
\newblock
\APACrefYearMonthDay{1994}{}{}.
\newblock
{\BBOQ}\APACrefatitle {A loop-top hard {X}-ray source in a compact solar flare
  as evidence for magnetic reconnection} {A loop-top hard {X}-ray source in a
  compact solar flare as evidence for magnetic reconnection}.{\BBCQ}
\newblock
\APACjournalVolNumPages{Nature}{371}{6497}{495--497}.
\newblock
\begin{APACrefDOI} \doi{10.1038/371495a0} \end{APACrefDOI}
\PrintBackRefs{\CurrentBib}

\bibitem [\protect \citeauthoryear {%
Mauk%
\ \protect \BOthers {.}}{%
Mauk%
\ \protect \BOthers {.}}{%
{\protect \APACyear {2016}}%
}]{%
mauk_energetic_2016}
\APACinsertmetastar {%
mauk_energetic_2016}%
\begin{APACrefauthors}%
Mauk, B\BPBI H.%
, Blake, J\BPBI B.%
, Baker, D\BPBI N.%
, Clemmons, J\BPBI H.%
, Reeves, G\BPBI D.%
, Spence, H\BPBI E.%
\BDBL {}Westlake, J\BPBI H.%
\end{APACrefauthors}%
\unskip\
\newblock
\APACrefYearMonthDay{2016}{}{}.
\newblock
{\BBOQ}\APACrefatitle {The {Energetic} {Particle} {Detector} ({EPD})
  {Investigation} and the {Energetic} {Ion} {Spectrometer} ({EIS}) for the
  {Magnetospheric} {Multiscale} ({MMS}) {Mission}} {The {Energetic} {Particle}
  {Detector} ({EPD}) {Investigation} and the {Energetic} {Ion} {Spectrometer}
  ({EIS}) for the {Magnetospheric} {Multiscale} ({MMS}) {Mission}}.{\BBCQ}
\newblock
\APACjournalVolNumPages{Space Science Reviews}{199}{1-4}{471--514}.
\newblock
\begin{APACrefDOI} \doi{10.1007/s11214-014-0055-5} \end{APACrefDOI}
\PrintBackRefs{\CurrentBib}

\bibitem [\protect \citeauthoryear {%
Merkin%
, Panov%
, Sorathia%
\BCBL {}\ \BBA {} Ukhorskiy%
}{%
Merkin%
\ \protect \BOthers {.}}{%
{\protect \APACyear {2019}}%
}]{%
merkin_contribution_2019}
\APACinsertmetastar {%
merkin_contribution_2019}%
\begin{APACrefauthors}%
Merkin, V\BPBI G.%
, Panov, E\BPBI V.%
, Sorathia, K\BPBI A.%
\BCBL {}\ \BBA {} Ukhorskiy, A\BPBI Y.%
\end{APACrefauthors}%
\unskip\
\newblock
\APACrefYearMonthDay{2019}{}{}.
\newblock
{\BBOQ}\APACrefatitle {Contribution of bursty bulk flows to the global
  dipolarization of the magnetotail during an isolated substorm} {Contribution
  of bursty bulk flows to the global dipolarization of the magnetotail during
  an isolated substorm}.{\BBCQ}
\newblock
\APACjournalVolNumPages{Journal of Geophysical Research: Space
  Physics}{124}{11}{8647--8668}.
\newblock
\begin{APACrefDOI} \doi{10.1029/2019JA026872} \end{APACrefDOI}
\PrintBackRefs{\CurrentBib}

\bibitem [\protect \citeauthoryear {%
Mitchell%
, Gkioulidou%
\BCBL {}\ \BBA {} Ukhorskiy%
}{%
Mitchell%
\ \protect \BOthers {.}}{%
{\protect \APACyear {2018}}%
}]{%
mitchell_energetic_2018}
\APACinsertmetastar {%
mitchell_energetic_2018}%
\begin{APACrefauthors}%
Mitchell, D\BPBI G.%
, Gkioulidou, M.%
\BCBL {}\ \BBA {} Ukhorskiy, A\BPBI Y.%
\end{APACrefauthors}%
\unskip\
\newblock
\APACrefYearMonthDay{2018}{}{}.
\newblock
{\BBOQ}\APACrefatitle {Energetic ion injections inside geosynchronous orbit:
  {Convection}- and drift-dominated, charge-dependent adiabatic energization
  ({W} = {qEd})} {Energetic ion injections inside geosynchronous orbit:
  {Convection}- and drift-dominated, charge-dependent adiabatic energization
  ({W} = {qEd})}.{\BBCQ}
\newblock
\APACjournalVolNumPages{Journal of Geophysical Research: Space
  Physics}{123}{8}{6360--6382}.
\newblock
\begin{APACrefDOI} \doi{10.1029/2018JA025556} \end{APACrefDOI}
\PrintBackRefs{\CurrentBib}

\bibitem [\protect \citeauthoryear {%
Nakamura%
\ \protect \BOthers {.}}{%
Nakamura%
\ \protect \BOthers {.}}{%
{\protect \APACyear {2002}}%
}]{%
nakamura_motion_2002}
\APACinsertmetastar {%
nakamura_motion_2002}%
\begin{APACrefauthors}%
Nakamura, R.%
, Baumjohann, W.%
, Klecker, B.%
, Bogdanova, Y.%
, Balogh, A.%
, Rème, H.%
\BDBL {}Runov, A.%
\end{APACrefauthors}%
\unskip\
\newblock
\APACrefYearMonthDay{2002}{}{}.
\newblock
{\BBOQ}\APACrefatitle {Motion of the dipolarization front during a flow burst
  event observed by {Cluster}} {Motion of the dipolarization front during a
  flow burst event observed by {Cluster}}.{\BBCQ}
\newblock
\APACjournalVolNumPages{Geophysical Research Letters}{29}{20}{1942}.
\newblock
\begin{APACrefDOI} \doi{10.1029/2002GL015763} \end{APACrefDOI}
\PrintBackRefs{\CurrentBib}

\bibitem [\protect \citeauthoryear {%
Nakamura%
\ \protect \BOthers {.}}{%
Nakamura%
\ \protect \BOthers {.}}{%
{\protect \APACyear {2004}}%
}]{%
nakamura_spatial_2004}
\APACinsertmetastar {%
nakamura_spatial_2004}%
\begin{APACrefauthors}%
Nakamura, R.%
, Baumjohann, W.%
, Mouikis, C.%
, Kistler, L\BPBI M.%
, Runov, A.%
, Volwerk, M.%
\BDBL {}Balogh, A.%
\end{APACrefauthors}%
\unskip\
\newblock
\APACrefYearMonthDay{2004}{}{}.
\newblock
{\BBOQ}\APACrefatitle {Spatial scale of high-speed flows in the plasma sheet
  observed by {Cluster}} {Spatial scale of high-speed flows in the plasma sheet
  observed by {Cluster}}.{\BBCQ}
\newblock
\APACjournalVolNumPages{Geophysical Research Letters}{31}{}{L09804}.
\newblock
\begin{APACrefDOI} \doi{10.1029/2004GL019558} \end{APACrefDOI}
\PrintBackRefs{\CurrentBib}

\bibitem [\protect \citeauthoryear {%
Ohtani%
}{%
Ohtani%
}{%
{\protect \APACyear {2004}}%
}]{%
ohtani_temporal_2004}
\APACinsertmetastar {%
ohtani_temporal_2004}%
\begin{APACrefauthors}%
Ohtani, S.%
\end{APACrefauthors}%
\unskip\
\newblock
\APACrefYearMonthDay{2004}{}{}.
\newblock
{\BBOQ}\APACrefatitle {Temporal structure of the fast convective flow in the
  plasma sheet: {Comparison} between observations and two-fluid simulations}
  {Temporal structure of the fast convective flow in the plasma sheet:
  {Comparison} between observations and two-fluid simulations}.{\BBCQ}
\newblock
\APACjournalVolNumPages{Journal of Geophysical Research}{109}{}{A03210}.
\newblock
\begin{APACrefDOI} \doi{10.1029/2003JA010002} \end{APACrefDOI}
\PrintBackRefs{\CurrentBib}

\bibitem [\protect \citeauthoryear {%
Panov%
\ \protect \BOthers {.}}{%
Panov%
\ \protect \BOthers {.}}{%
{\protect \APACyear {2010}}%
}]{%
panov_multiple_2010}
\APACinsertmetastar {%
panov_multiple_2010}%
\begin{APACrefauthors}%
Panov, E\BPBI V.%
, Nakamura, R.%
, Baumjohann, W.%
, Angelopoulos, V.%
, Petrukovich, A\BPBI A.%
, Retinò, A.%
\BDBL {}Larson, D.%
\end{APACrefauthors}%
\unskip\
\newblock
\APACrefYearMonthDay{2010}{}{}.
\newblock
{\BBOQ}\APACrefatitle {Multiple overshoot and rebound of a bursty bulk flow}
  {Multiple overshoot and rebound of a bursty bulk flow}.{\BBCQ}
\newblock
\APACjournalVolNumPages{Geophysical Research Letters}{37}{8}{}.
\newblock
\begin{APACrefDOI} \doi{10.1029/2009GL041971} \end{APACrefDOI}
\PrintBackRefs{\CurrentBib}

\bibitem [\protect \citeauthoryear {%
Phan%
\ \protect \BOthers {.}}{%
Phan%
\ \protect \BOthers {.}}{%
{\protect \APACyear {2000}}%
}]{%
phan_extended_2000}
\APACinsertmetastar {%
phan_extended_2000}%
\begin{APACrefauthors}%
Phan, T\BPBI D.%
, Kistler, L\BPBI M.%
, Klecker, B.%
, Haerendel, G.%
, Paschmann, G.%
, Sonnerup, B\BPBI U\BPBI {\"O}.%
\BDBL {}Reme, H.%
\end{APACrefauthors}%
\unskip\
\newblock
\APACrefYearMonthDay{2000}{}{}.
\newblock
{\BBOQ}\APACrefatitle {Extended magnetic reconnection at the {Earth}'s
  magnetopause from detection of bi-directional jets} {Extended magnetic
  reconnection at the {Earth}'s magnetopause from detection of bi-directional
  jets}.{\BBCQ}
\newblock
\APACjournalVolNumPages{Nature}{404}{6780}{848--850}.
\newblock
\begin{APACrefDOI} \doi{10.1038/35009050} \end{APACrefDOI}
\PrintBackRefs{\CurrentBib}

\bibitem [\protect \citeauthoryear {%
Pollock%
\ \protect \BOthers {.}}{%
Pollock%
\ \protect \BOthers {.}}{%
{\protect \APACyear {2016}}%
}]{%
pollock_fast_2016}
\APACinsertmetastar {%
pollock_fast_2016}%
\begin{APACrefauthors}%
Pollock, C.%
, Moore, T.%
, Jacques, A.%
, Burch, J.%
, Gliese, U.%
, Saito, Y.%
\BDBL {}Zeuch, M.%
\end{APACrefauthors}%
\unskip\
\newblock
\APACrefYearMonthDay{2016}{}{}.
\newblock
{\BBOQ}\APACrefatitle {Fast {Plasma} {Investigation} for {Magnetospheric}
  {Multiscale}} {Fast {Plasma} {Investigation} for {Magnetospheric}
  {Multiscale}}.{\BBCQ}
\newblock
\APACjournalVolNumPages{Space Science Reviews}{199}{1-4}{331--406}.
\newblock
\begin{APACrefDOI} \doi{10.1007/s11214-016-0245-4} \end{APACrefDOI}
\PrintBackRefs{\CurrentBib}

\bibitem [\protect \citeauthoryear {%
Pritchett%
\ \BBA {} Coroniti%
}{%
Pritchett%
\ \BBA {} Coroniti%
}{%
{\protect \APACyear {2010}}%
}]{%
pritchett_kinetic_2010}
\APACinsertmetastar {%
pritchett_kinetic_2010}%
\begin{APACrefauthors}%
Pritchett, P\BPBI L.%
\BCBT {}\ \BBA {} Coroniti, F\BPBI V.%
\end{APACrefauthors}%
\unskip\
\newblock
\APACrefYearMonthDay{2010}{}{}.
\newblock
{\BBOQ}\APACrefatitle {A kinetic ballooning/interchange instability in the
  magnetotail} {A kinetic ballooning/interchange instability in the
  magnetotail}.{\BBCQ}
\newblock
\APACjournalVolNumPages{Journal of Geophysical Research: Space
  Physics}{115}{A06301}{}.
\newblock
\begin{APACrefDOI} \doi{10.1029/2009JA014752} \end{APACrefDOI}
\PrintBackRefs{\CurrentBib}

\bibitem [\protect \citeauthoryear {%
Pudritz%
, Hardcastle%
\BCBL {}\ \BBA {} Gabuzda%
}{%
Pudritz%
\ \protect \BOthers {.}}{%
{\protect \APACyear {2012}}%
}]{%
pudritz_magnetic_2012}
\APACinsertmetastar {%
pudritz_magnetic_2012}%
\begin{APACrefauthors}%
Pudritz, R\BPBI E.%
, Hardcastle, M\BPBI J.%
\BCBL {}\ \BBA {} Gabuzda, D\BPBI C.%
\end{APACrefauthors}%
\unskip\
\newblock
\APACrefYearMonthDay{2012}{}{}.
\newblock
{\BBOQ}\APACrefatitle {Magnetic fields in astrophysical jets: {From} launch to
  termination} {Magnetic fields in astrophysical jets: {From} launch to
  termination}.{\BBCQ}
\newblock
\APACjournalVolNumPages{Space Science Reviews}{169}{1-4}{27--72}.
\newblock
\begin{APACrefDOI} \doi{10.1007/s11214-012-9895-z} \end{APACrefDOI}
\PrintBackRefs{\CurrentBib}

\bibitem [\protect \citeauthoryear {%
Runov%
\ \protect \BOthers {.}}{%
Runov%
\ \protect \BOthers {.}}{%
{\protect \APACyear {2015}}%
}]{%
runov_average_2015}
\APACinsertmetastar {%
runov_average_2015}%
\begin{APACrefauthors}%
Runov, A.%
, Angelopoulos, V.%
, Gabrielse, C.%
, Liu, J.%
, Turner, D\BPBI L.%
\BCBL {}\ \BBA {} Zhou, X.%
\end{APACrefauthors}%
\unskip\
\newblock
\APACrefYearMonthDay{2015}{}{}.
\newblock
{\BBOQ}\APACrefatitle {Average thermodynamic and spectral properties of plasma
  in and around dipolarizing flux bundles} {Average thermodynamic and spectral
  properties of plasma in and around dipolarizing flux bundles}.{\BBCQ}
\newblock
\APACjournalVolNumPages{Journal of Geophysical Research: Space
  Physics}{120}{6}{4369--4383}.
\newblock
\begin{APACrefDOI} \doi{10.1002/2015JA021166} \end{APACrefDOI}
\PrintBackRefs{\CurrentBib}

\bibitem [\protect \citeauthoryear {%
Runov%
\ \protect \BOthers {.}}{%
Runov%
\ \protect \BOthers {.}}{%
{\protect \APACyear {2009}}%
}]{%
runov_themis_2009}
\APACinsertmetastar {%
runov_themis_2009}%
\begin{APACrefauthors}%
Runov, A.%
, Angelopoulos, V.%
, Sitnov, M\BPBI I.%
, Sergeev, V\BPBI A.%
, Bonnell, J.%
, McFadden, J\BPBI P.%
\BDBL {}Auster, U.%
\end{APACrefauthors}%
\unskip\
\newblock
\APACrefYearMonthDay{2009}{}{}.
\newblock
{\BBOQ}\APACrefatitle {{THEMIS} observations of an earthward-propagating
  dipolarization front} {{THEMIS} observations of an earthward-propagating
  dipolarization front}.{\BBCQ}
\newblock
\APACjournalVolNumPages{Geophysical Research Letters}{36}{}{L14106}.
\newblock
\begin{APACrefDOI} \doi{10.1029/2009GL038980} \end{APACrefDOI}
\PrintBackRefs{\CurrentBib}

\bibitem [\protect \citeauthoryear {%
Runov%
\ \protect \BOthers {.}}{%
Runov%
\ \protect \BOthers {.}}{%
{\protect \APACyear {2011}}%
}]{%
runov_themis_2011}
\APACinsertmetastar {%
runov_themis_2011}%
\begin{APACrefauthors}%
Runov, A.%
, Angelopoulos, V.%
, Zhou, X\BHBI Z.%
, Zhang, X\BHBI J.%
, Li, S.%
, Plaschke, F.%
\BCBL {}\ \BBA {} Bonnell, J.%
\end{APACrefauthors}%
\unskip\
\newblock
\APACrefYearMonthDay{2011}{}{}.
\newblock
{\BBOQ}\APACrefatitle {A {THEMIS} multicase study of dipolarization fronts in
  the magnetotail plasma sheet} {A {THEMIS} multicase study of dipolarization
  fronts in the magnetotail plasma sheet}.{\BBCQ}
\newblock
\APACjournalVolNumPages{Journal of Geophysical Research: Space
  Physics}{116}{}{A05216}.
\newblock
\begin{APACrefDOI} \doi{10.1029/2010JA016316} \end{APACrefDOI}
\PrintBackRefs{\CurrentBib}

\bibitem [\protect \citeauthoryear {%
Russell%
\ \protect \BOthers {.}}{%
Russell%
\ \protect \BOthers {.}}{%
{\protect \APACyear {2016}}%
}]{%
russell_magnetospheric_2016}
\APACinsertmetastar {%
russell_magnetospheric_2016}%
\begin{APACrefauthors}%
Russell, C\BPBI T.%
, Anderson, B\BPBI J.%
, Baumjohann, W.%
, Bromund, K\BPBI R.%
, Dearborn, D.%
, Fischer, D.%
\BDBL {}Richter, I.%
\end{APACrefauthors}%
\unskip\
\newblock
\APACrefYearMonthDay{2016}{}{}.
\newblock
{\BBOQ}\APACrefatitle {The {Magnetospheric} {Multiscale} {Magnetometers}} {The
  {Magnetospheric} {Multiscale} {Magnetometers}}.{\BBCQ}
\newblock
\APACjournalVolNumPages{Space Science Reviews}{199}{1-4}{189--256}.
\newblock
\begin{APACrefDOI} \doi{10.1007/s11214-014-0057-3} \end{APACrefDOI}
\PrintBackRefs{\CurrentBib}

\bibitem [\protect \citeauthoryear {%
Schulz%
\ \BBA {} Lanzerotti%
}{%
Schulz%
\ \BBA {} Lanzerotti%
}{%
{\protect \APACyear {1974}}%
}]{%
schulz_particle_1974}
\APACinsertmetastar {%
schulz_particle_1974}%
\begin{APACrefauthors}%
Schulz, M.%
\BCBT {}\ \BBA {} Lanzerotti, L\BPBI J.%
\end{APACrefauthors}%
\unskip\
\newblock
\APACrefYear{1974}.
\newblock
\APACrefbtitle {Particle {Diffusion} in the {Radiation} {Belts}} {Particle
  {Diffusion} in the {Radiation} {Belts}}\ (\BVOL~7; J\BPBI G.~Roederer,
  \BED{}).
\newblock
\APACaddressPublisher{Berlin, Heidelberg}{Springer Berlin Heidelberg}.
\newblock
\begin{APACrefDOI} \doi{10.1007/978-3-642-65675-0} \end{APACrefDOI}
\PrintBackRefs{\CurrentBib}

\bibitem [\protect \citeauthoryear {%
Sergeev%
\ \protect \BOthers {.}}{%
Sergeev%
\ \protect \BOthers {.}}{%
{\protect \APACyear {2009}}%
}]{%
sergeev_kinetic_2009}
\APACinsertmetastar {%
sergeev_kinetic_2009}%
\begin{APACrefauthors}%
Sergeev, V.%
, Angelopoulos, V.%
, Apatenkov, S.%
, Bonnell, J.%
, Ergun, R.%
, Nakamura, R.%
\BDBL {}Runov, A.%
\end{APACrefauthors}%
\unskip\
\newblock
\APACrefYearMonthDay{2009}{}{}.
\newblock
{\BBOQ}\APACrefatitle {Kinetic structure of the sharp injection/dipolarization
  front in the flow-braking region} {Kinetic structure of the sharp
  injection/dipolarization front in the flow-braking region}.{\BBCQ}
\newblock
\APACjournalVolNumPages{Geophysical Research Letters}{36}{L21105}{}.
\newblock
\begin{APACrefDOI} \doi{10.1029/2009GL040658} \end{APACrefDOI}
\PrintBackRefs{\CurrentBib}

\bibitem [\protect \citeauthoryear {%
Sitnov%
\ \protect \BOthers {.}}{%
Sitnov%
\ \protect \BOthers {.}}{%
{\protect \APACyear {2019}}%
}]{%
sitnov_explosive_2019}
\APACinsertmetastar {%
sitnov_explosive_2019}%
\begin{APACrefauthors}%
Sitnov, M\BPBI I.%
, Birn, J.%
, Ferdousi, B.%
, Gordeev, E.%
, Khotyaintsev, Y\BPBI V.%
, Merkin, V\BPBI G.%
\BDBL {}Zhou, X.%
\end{APACrefauthors}%
\unskip\
\newblock
\APACrefYearMonthDay{2019}{}{}.
\newblock
{\BBOQ}\APACrefatitle {Explosive magnetotail activity} {Explosive magnetotail
  activity}.{\BBCQ}
\newblock
\APACjournalVolNumPages{Space Science Reviews}{215}{4}{31}.
\newblock
\begin{APACrefDOI} \doi{10.1007/s11214-019-0599-5} \end{APACrefDOI}
\PrintBackRefs{\CurrentBib}

\bibitem [\protect \citeauthoryear {%
Sitnov%
, Swisdak%
\BCBL {}\ \BBA {} Divin%
}{%
Sitnov%
\ \protect \BOthers {.}}{%
{\protect \APACyear {2009}}%
}]{%
sitnov_dipolarization_2009}
\APACinsertmetastar {%
sitnov_dipolarization_2009}%
\begin{APACrefauthors}%
Sitnov, M\BPBI I.%
, Swisdak, M.%
\BCBL {}\ \BBA {} Divin, A\BPBI V.%
\end{APACrefauthors}%
\unskip\
\newblock
\APACrefYearMonthDay{2009}{}{}.
\newblock
{\BBOQ}\APACrefatitle {Dipolarization fronts as a signature of transient
  reconnection in the magnetotail} {Dipolarization fronts as a signature of
  transient reconnection in the magnetotail}.{\BBCQ}
\newblock
\APACjournalVolNumPages{Journal of Geophysical Research: Space
  Physics}{114}{A04202}{}.
\newblock
\begin{APACrefDOI} \doi{10.1029/2008JA013980} \end{APACrefDOI}
\PrintBackRefs{\CurrentBib}

\bibitem [\protect \citeauthoryear {%
Tsyganenko%
}{%
Tsyganenko%
}{%
{\protect \APACyear {1989}}%
}]{%
tsyganenko_magnetospheric_1989}
\APACinsertmetastar {%
tsyganenko_magnetospheric_1989}%
\begin{APACrefauthors}%
Tsyganenko, N.%
\end{APACrefauthors}%
\unskip\
\newblock
\APACrefYearMonthDay{1989}{}{}.
\newblock
{\BBOQ}\APACrefatitle {A magnetospheric magnetic field model with a warped tail
  current sheet} {A magnetospheric magnetic field model with a warped tail
  current sheet}.{\BBCQ}
\newblock
\APACjournalVolNumPages{Planetary and Space Science}{37}{1}{5--20}.
\newblock
\begin{APACrefDOI} \doi{10.1016/0032-0633(89)90066-4} \end{APACrefDOI}
\PrintBackRefs{\CurrentBib}

\bibitem [\protect \citeauthoryear {%
Turner%
\ \protect \BOthers {.}}{%
Turner%
\ \protect \BOthers {.}}{%
{\protect \APACyear {2016}}%
}]{%
turner_energy_2016}
\APACinsertmetastar {%
turner_energy_2016}%
\begin{APACrefauthors}%
Turner, D\BPBI L.%
, Fennell, J\BPBI F.%
, Blake, J\BPBI B.%
, Clemmons, J\BPBI H.%
, Mauk, B\BPBI H.%
, Cohen, I\BPBI J.%
\BDBL {}Burch, J\BPBI L.%
\end{APACrefauthors}%
\unskip\
\newblock
\APACrefYearMonthDay{2016}{}{}.
\newblock
{\BBOQ}\APACrefatitle {Energy limits of electron acceleration in the plasma
  sheet during substorms: {A} case study with the {Magnetospheric} {Multiscale}
  ({MMS}) mission} {Energy limits of electron acceleration in the plasma sheet
  during substorms: {A} case study with the {Magnetospheric} {Multiscale}
  ({MMS}) mission}.{\BBCQ}
\newblock
\APACjournalVolNumPages{Geophysical Research Letters}{43}{}{7785--7794}.
\newblock
\begin{APACrefDOI} \doi{10.1002/2016GL069691} \end{APACrefDOI}
\PrintBackRefs{\CurrentBib}

\bibitem [\protect \citeauthoryear {%
Ukhorskiy%
, Sitnov%
, Merkin%
\BCBL {}\ \BBA {} Artemyev%
}{%
Ukhorskiy%
\ \protect \BOthers {.}}{%
{\protect \APACyear {2013}}%
}]{%
ukhorskiy_rapid_2013}
\APACinsertmetastar {%
ukhorskiy_rapid_2013}%
\begin{APACrefauthors}%
Ukhorskiy, A\BPBI Y.%
, Sitnov, M\BPBI I.%
, Merkin, V\BPBI G.%
\BCBL {}\ \BBA {} Artemyev, A\BPBI V.%
\end{APACrefauthors}%
\unskip\
\newblock
\APACrefYearMonthDay{2013}{}{}.
\newblock
{\BBOQ}\APACrefatitle {Rapid acceleration of protons upstream of earthward
  propagating dipolarization fronts} {Rapid acceleration of protons upstream of
  earthward propagating dipolarization fronts}.{\BBCQ}
\newblock
\APACjournalVolNumPages{Journal of Geophysical Research: Space
  Physics}{118}{8}{4952--4962}.
\newblock
\begin{APACrefDOI} \doi{10.1002/jgra.50452} \end{APACrefDOI}
\PrintBackRefs{\CurrentBib}

\bibitem [\protect \citeauthoryear {%
Ukhorskiy%
, Sitnov%
, Merkin%
, Gkioulidou%
\BCBL {}\ \BBA {} Mitchell%
}{%
Ukhorskiy%
\ \protect \BOthers {.}}{%
{\protect \APACyear {2017}}%
}]{%
ukhorskiy_ion_2017}
\APACinsertmetastar {%
ukhorskiy_ion_2017}%
\begin{APACrefauthors}%
Ukhorskiy, A\BPBI Y.%
, Sitnov, M\BPBI I.%
, Merkin, V\BPBI G.%
, Gkioulidou, M.%
\BCBL {}\ \BBA {} Mitchell, D\BPBI G.%
\end{APACrefauthors}%
\unskip\
\newblock
\APACrefYearMonthDay{2017}{}{}.
\newblock
{\BBOQ}\APACrefatitle {Ion acceleration at dipolarization fronts in the inner
  magnetosphere} {Ion acceleration at dipolarization fronts in the inner
  magnetosphere}.{\BBCQ}
\newblock
\APACjournalVolNumPages{Journal of Geophysical Research: Space
  Physics}{122}{3}{3040--3054}.
\newblock
\begin{APACrefDOI} \doi{10.1002/2016JA023304} \end{APACrefDOI}
\PrintBackRefs{\CurrentBib}

\bibitem [\protect \citeauthoryear {%
Ukhorskiy%
\ \protect \BOthers {.}}{%
Ukhorskiy%
\ \protect \BOthers {.}}{%
{\protect \APACyear {2018}}%
}]{%
ukhorskiy_ion_2018}
\APACinsertmetastar {%
ukhorskiy_ion_2018}%
\begin{APACrefauthors}%
Ukhorskiy, A\BPBI Y.%
, Sorathia, K\BPBI A.%
, Merkin, V\BPBI G.%
, Sitnov, M\BPBI I.%
, Mitchell, D\BPBI G.%
\BCBL {}\ \BBA {} Gkioulidou, M.%
\end{APACrefauthors}%
\unskip\
\newblock
\APACrefYearMonthDay{2018}{}{}.
\newblock
{\BBOQ}\APACrefatitle {Ion trapping and acceleration at dipolarization fronts:
  {High}-resolution {MHD} and test-particle simulations} {Ion trapping and
  acceleration at dipolarization fronts: {High}-resolution {MHD} and
  test-particle simulations}.{\BBCQ}
\newblock
\APACjournalVolNumPages{Journal of Geophysical Research: Space
  Physics}{123}{7}{5580--5589}.
\newblock
\begin{APACrefDOI} \doi{10.1029/2018JA025370} \end{APACrefDOI}
\PrintBackRefs{\CurrentBib}

\bibitem [\protect \citeauthoryear {%
Young%
\ \protect \BOthers {.}}{%
Young%
\ \protect \BOthers {.}}{%
{\protect \APACyear {2016}}%
}]{%
young_hot_2016}
\APACinsertmetastar {%
young_hot_2016}%
\begin{APACrefauthors}%
Young, D\BPBI T.%
, Burch, J\BPBI L.%
, Gomez, R\BPBI G.%
, De~Los~Santos, A.%
, Miller, G\BPBI P.%
, Wilson, P.%
\BDBL {}Webster, J\BPBI M.%
\end{APACrefauthors}%
\unskip\
\newblock
\APACrefYearMonthDay{2016}{}{}.
\newblock
{\BBOQ}\APACrefatitle {Hot {Plasma} {Composition} {Analyzer} for the
  {Magnetospheric} {Multiscale} {Mission}} {Hot {Plasma} {Composition}
  {Analyzer} for the {Magnetospheric} {Multiscale} {Mission}}.{\BBCQ}
\newblock
\APACjournalVolNumPages{Space Science Reviews}{199}{1-4}{407--470}.
\newblock
\begin{APACrefDOI} \doi{10.1007/s11214-014-0119-6} \end{APACrefDOI}
\PrintBackRefs{\CurrentBib}

\bibitem [\protect \citeauthoryear {%
Zhou%
, Angelopoulos%
, Sergeev%
\BCBL {}\ \BBA {} Runov%
}{%
Zhou%
\ \protect \BOthers {.}}{%
{\protect \APACyear {2010}}%
}]{%
zhou_accelerated_2010}
\APACinsertmetastar {%
zhou_accelerated_2010}%
\begin{APACrefauthors}%
Zhou, X\BHBI Z.%
, Angelopoulos, V.%
, Sergeev, V\BPBI A.%
\BCBL {}\ \BBA {} Runov, A.%
\end{APACrefauthors}%
\unskip\
\newblock
\APACrefYearMonthDay{2010}{}{}.
\newblock
{\BBOQ}\APACrefatitle {Accelerated ions ahead of earthward propagating
  dipolarization fronts} {Accelerated ions ahead of earthward propagating
  dipolarization fronts}.{\BBCQ}
\newblock
\APACjournalVolNumPages{Journal of Geophysical Research: Space
  Physics}{115}{A00I03}{}.
\newblock
\begin{APACrefDOI} \doi{10.1029/2010JA015481} \end{APACrefDOI}
\PrintBackRefs{\CurrentBib}

\bibitem [\protect \citeauthoryear {%
Zhou%
, Angelopoulos%
, Sergeev%
\BCBL {}\ \BBA {} Runov%
}{%
Zhou%
\ \protect \BOthers {.}}{%
{\protect \APACyear {2011}}%
}]{%
zhou_nature_2011}
\APACinsertmetastar {%
zhou_nature_2011}%
\begin{APACrefauthors}%
Zhou, X\BHBI Z.%
, Angelopoulos, V.%
, Sergeev, V\BPBI A.%
\BCBL {}\ \BBA {} Runov, A.%
\end{APACrefauthors}%
\unskip\
\newblock
\APACrefYearMonthDay{2011}{}{}.
\newblock
{\BBOQ}\APACrefatitle {On the nature of precursor flows upstream of advancing
  dipolarization fronts} {On the nature of precursor flows upstream of
  advancing dipolarization fronts}.{\BBCQ}
\newblock
\APACjournalVolNumPages{Journal of Geophysical Research: Space
  Physics}{116}{A03222}{}.
\newblock
\begin{APACrefDOI} \doi{10.1029/2010JA016165} \end{APACrefDOI}
\PrintBackRefs{\CurrentBib}

\end{thebibliography}


\end{document}